**Correspondence to:**
A. Broberg Skeltved,
abskeltved@yahoo.no

# Modeling the relativistic runaway electron avalanche and the feedback mechanism with GEANT4

**Alexander Broberg Skeltved[1], Nikolai Østgaard[1], Brant Carlson[1,2], Thomas Gjesteland[1], and Sebastien Celestin[3]**

[1]Birkeland Centre for Space Science, Institute of Physics and Technology, University of Bergen, Bergen, Norway, [2]Physics and Astronomy, Carthage College, Kenosha, Wisconsin, USA, [3]Laboratory of Physics and Chemistry of the Environment and Space, University of Orleans, CNRS, Orleans, France

**Abstract** This paper presents the first study that uses the GEometry ANd Tracking 4 (GEANT4) toolkit to do quantitative comparisons with other modeling results related to the production of terrestrial gamma ray flashes and high-energy particle emission from thunderstorms. We will study the relativistic runaway electron avalanche (RREA) and the relativistic feedback process, as well as the production of bremsstrahlung photons from runaway electrons. The Monte Carlo simulations take into account the effects of electron ionization, electron by electron (MÃller), and electron by positron (Bhabha) scattering as well as the bremsstrahlung process and pair production, in the 250 eV to 100 GeV energy range. Our results indicate that the multiplication of electrons during the development of RREAs and under the influence of feedback are consistent with previous estimates. This is important to validate GEANT4 as a tool to model RREAs and feedback in homogeneous electric fields. We also determine the ratio of bremsstrahlung photons to energetic electrons $N_\gamma/N_e$. We then show that the ratio has a dependence on the electric field, which can be expressed by the avalanche time $\tau(E)$ and the bremsstrahlung coefficient $\alpha(\varepsilon)$. In addition, we present comparisons of GEANT4 simulations performed with a "standard" and a "low-energy" physics list both validated in the 1 keV to 100 GeV energy range. This comparison shows that the choice of physics list used in GEANT4 simulations has a significant effect on the results.

## 1. Introduction

Terrestrial gamma ray flashes were first discovered in the early 1990 by the Burst And Transient Source Experiment (BATSE) on board NASA's Compton Gamma Ray Observatory [*Fishman et al.*, 1994]. Since then, the observations of these submillisecond bursts of up to several tens of MeV photons have been confirmed in multiple studies [*Smith et al.*, 2005; *Briggs et al.*, 2010; *Marisaldi et al.*, 2010]. From modeling results and comparisons with the average photon energy spectrums obtained by satellite measurements, terrestrial gamma ray flashes (TGF) production has been determined to occur below 21 km altitude inside thundercloud regions [*Dwyer and Smith*, 2005; *Carlson et al.*, 2007; *Østgaard et al.*, 2008; *Gjesteland et al.*, 2010]. Measurements have shown that the intensities of TGFs range from $10^{14}$ photons [*Østgaard et al.*, 2012] to $10^{17}$ [*Dwyer and Smith*, 2005] if they are produced at 15 km altitude. *Hansen et al.* [2013] show that this intensity may vary with up to 3 orders of magnitude depending on the production altitude assumed. *Østgaard et al.* [2012], from the measurements available so far, argue that it cannot be ruled out that all discharges produce TGFs and that the lower intensity limit is then $10^{12}$, again given a production altitude of 15 km. The number of electrons that are required, at source altitude, to produce these large fluxes of photons is generally assumed to be between the same and 1 order of magnitude larger than the number of photons.

The exact mechanism responsible for the production and multiplication of the energetic electrons is not yet fully understood. It is known, however, that the electric fields generated inside thunderclouds are capable of accelerating electrons to the energies required. Two leading theories currently exist to explain the multiplication of the energetic electrons and the subsequent production of bremsstrahlung photons.

1. The thermal acceleration of electrons in the tips of streamers and the subsequent acceleration during the stepping of lightning leaders [*Moss et al.*, 2006; *Williams et al.*, 2006; *Dwyer*, 2008, *Carlson et al.*, 2009, 2010; *Chanrion and Neubert*, 2010; *Celestin and Pasko*, 2011; *Xu et al.*, 2012].
2. The initiation of high-energy electrons from seed particles such as the products of cosmic rays. The continued multiplication and acceleration of these electrons through the relativistic runaway electron avalanche (RREA) process in the ambient electric field of the thundercloud [*Gurevich et al.*, 1992]. Finally, further multiplication of RREAs from backscattering photons and positrons, which is the feedback mechanism [*Dwyer*, 2003, 2007; *Dwyer and Babich*, 2011; *Dwyer*, 2012].

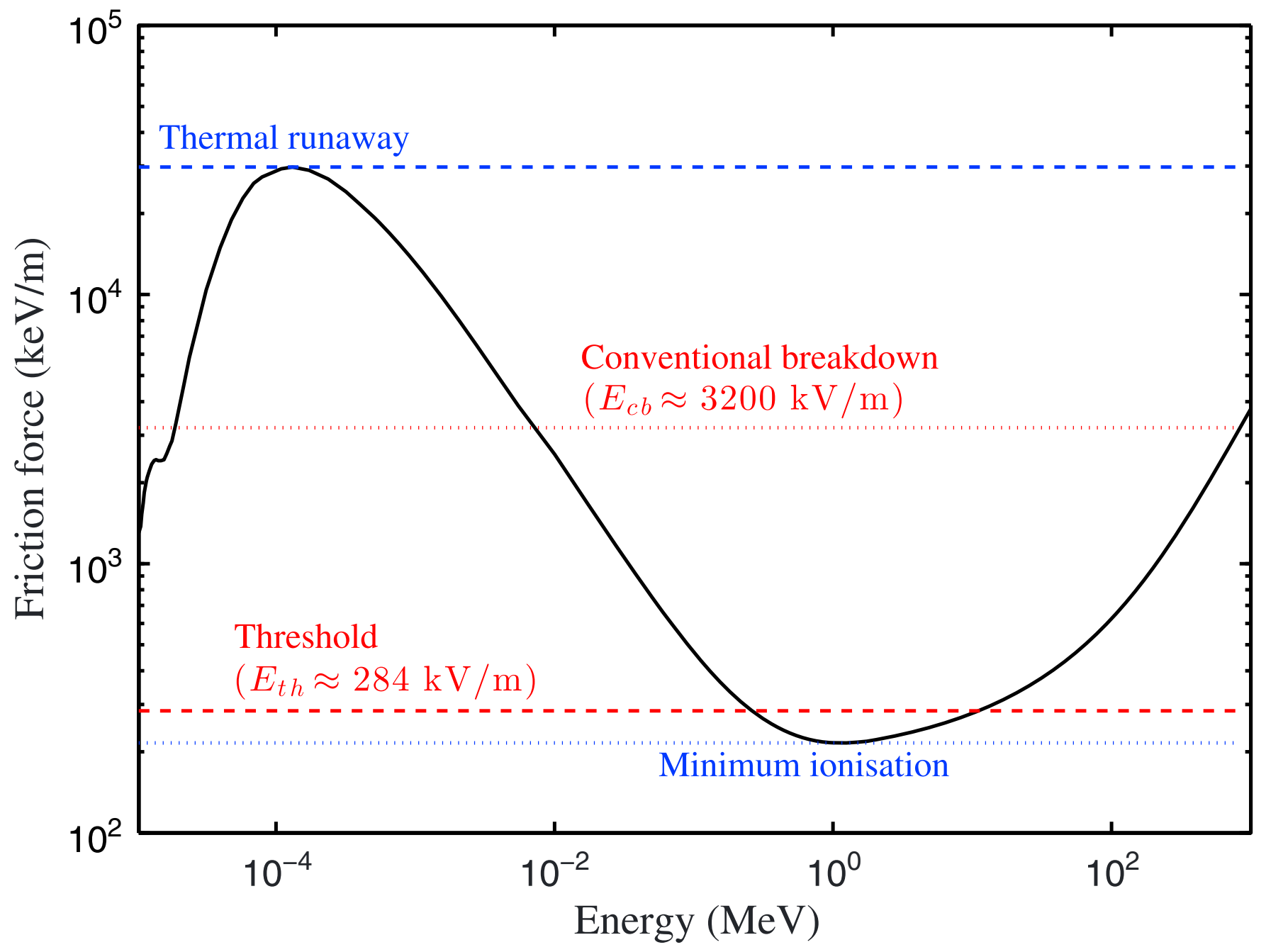

**Figure 1.** The friction force experienced by electrons in air at sea level with respect to their kinetic energy (solid black line). The dotted red line indicates the conventional breakdown field, $E_{cb}$ = 3200 kV/m. The dashed red line show the effective minimum threshold force experienced by runaway electrons and corresponds to $E_{th}$ = 284 kV/m [*Dwyer*, 2003]. The dashed blue line shows the upper threshold for thermal runaway to occur, and the lower dotted blue line indicates the minimum ionization threshold. The data set was obtained from *International Commission on Radiation Units and Measurements* [1984].

The electron multiplication through the feedback mechanism will be the subject of this paper. This mechanism is constrained by the available potential in the cloud as well as the strength, location, and vertical extent of the electric field. We will also discuss the ratio of bremsstrahlung photons to electrons at the end of the electric field region. This ratio is important when estimating the total number of electrons that is required at the source of TGFs. Results will be obtained using the GEometry And Tracking 4 (GEANT4, version 9.6) programming toolkit, which will be discussed in section 2.

This study completes the work first presented at the European Geophysical Union 2013 spring meeting [*Skeltved et al.*, 2013]. Another study that uses GEANT4 simulations of relativistic feedback discharges was presented at the American Geophysical Union 2013 fall meeting [*Gwen et al.*, 2013].

### 1.1. Runaway Electrons

*Wilson* [1925], based on observations of the tracks of energetic electrons in a cloud chamber, suggested a theory to explain the behavior of energetic electrons in a thundercloud. Wilson proposed that energetic electrons in air, such as can be produced from cosmic rays, can be accelerated to large energies by the strong electric fields produced in thunderclouds. These electric fields must be sufficiently strong to oppose the effective friction force resulting from electron interactions with air molecules. Electrons that continue to be accelerated then become runaway electrons (REs). The effective frictional force in air at sea level pressure and density, with respect to the kinetic energy of the electron, is shown in Figure 1. The minimum friction force for REs is experienced by electrons with a kinetic energy of ≈1 MeV. Monte Carlo simulations show that in order for REs to propagate large distances, the electric field threshold $E_{th}$ is approximately 30% larger than the minimum ionization threshold and is equal to 284 kV/m (dashed red line) [*Dwyer*, 2003]. This is due to the effect of elastic scattering, which causes electrons to scatter out of alignment of the electric field. The upper limit, where local ionization can occur and which will cause streamers and subsequent lightning discharges to form, is called the conventional breakdown field, $E_{cb} \approx 3200$ kV/m (dotted red line). Thermal runaway occurs at approximately $10E_{cb}$. The average energy gained $d\varepsilon$ by runaway electrons traveling a given distance $dz$ through a thundercloud can be expressed as a function of the electric field $E$ and the opposing friction force $F_d$

$$d\varepsilon = dz(eE - F_d), \tag{1}$$

where $e$ is the elementary charge.

### 1.2. Relativistic Runaway Electron Avalanche

In 1984 *McCarthy and Parks* [1985] reported intensive bursts of X-rays, which lasted a few seconds each and emanated from regions inside thunderstorms. McCarthy and Parks suggested that runaway electrons, which Wilson first described, produced the measured bremsstrahlung X-rays. However, this process could not explain the measured fluxes by itself. *Gurevich et al.* [1992] then introduced the idea that runaway electrons could undergo a multiplication process through high-energy electron-electron elastic scattering (primarily Møller scattering), and form an avalanche process antiparallel to the electric field. This process is called a relativistic runaway electron avalanche (RREA). Wilson also appears to have been aware of this avalanche effect as he in his personal notes described it as the "Snowball effect" [*Williams*, 2010].

The initiation of RREAs still relies upon the presence of seed electrons. A suggested source of high-energy seed electrons in the Earth's atmosphere are the extensive air showers (EAS) resulting from cosmic rays. *Carlson et al.* [2008] calculated that cosmic ray secondaries will be present within $\approx 1\,\mu$s in spherical volumes of radius 100 m to 3 km at altitudes of 0.5 km to 29.5 km. We can then assume that RREAs will quickly be initiated when a region within a thundercloud is of sufficient electric field strength.

*Gurevich et al.* [1992] showed that the number of runaway electrons in one avalanche increases with time $t$ and distance $z$ and can be expressed as

$$\frac{dN_{\mathrm{RREA}}}{dz} = \frac{1}{\lambda} N_{\mathrm{RREA}}, \tag{2}$$

$$\frac{dN_{\mathrm{RREA}}}{dt} = \frac{1}{\tau} N_{\mathrm{RREA}}, \tag{3}$$

where $\lambda$ is the avalanche growth length and $\tau$ is the avalanche growth time. Integrating over the total length of the avalanche region, from $z = 0$ to $z = L$, we get the total number of runaway electrons produced in a RREA

$$N_{\mathrm{RREA}} = N_0 \exp(L/\lambda), \tag{4}$$

or, over total time, from $t = 0$ to $t = t$

$$N_{\mathrm{RREA}} = N_0 \exp(t/\tau). \tag{5}$$

Based on the MC model presented in *Dwyer* [2003], *Coleman and Dwyer* [2006] presented the *e*-folding length or avalanche length $\lambda$ expressed as a function of electric field strength, $E$,

$$\lambda(E) = \frac{7300\ \mathrm{kV}}{E - 276\ \mathrm{kV/m}}. \tag{6}$$

In addition, the avalanche time $\tau$ was expressed as

$$\tau(E) = \frac{27.3\ \mathrm{kV}\ \mu\mathrm{s/m}}{E - 277\ \mathrm{kV/m}} = \frac{\lambda}{0.89c}, \tag{7}$$

where $c$ is the speed of light and $0.89c$ is the average speed of the propagating avalanche (also determined from Monte Carlo (MC) simulations [*Coleman and Dwyer*, 2006]). The number of produced bremsstrahlung photons can then be determined by multiplying the number of electrons (equation (5)) by a factor $\alpha(\varepsilon_{\mathrm{th}})\tau(E)$:

$$N_\gamma = \alpha(\varepsilon_{\mathrm{th}})\tau(E) N_0 \exp(t/\tau(E)), \tag{8}$$

where we assume that $t >> \tau(E)$, $\alpha$ is the bremsstrahlung coefficient, $\tau(E)$ is the avalanche growth time of RREAs, and $N_0$ is the number of initial seed electrons (see Appendix B for a derivation of the bremsstrahlung photon to runaway electron ratio).

In a review of terrestrial gamma ray flashes, which includes comparisons of studies concerning RREAs, *Dwyer et al.* [2012] present the electron energy spectrum. If the initial number of cosmic ray seed electrons is $N_0$, the total number of electrons in the RREA at the end of the avalanche region is given by equation (4). We can then find the change $dN(z)$ over a distance $dz$, and by rewriting equation (1) to $dz = d\varepsilon/(eE - F_d)$, we then derive the electron energy distribution function (EEDF) after a few avalanche lengths, or the number of runaway electrons per unit energy,

$$f_{\mathrm{re}} = \frac{dN_{\mathrm{RREA}}(\varepsilon)}{d\varepsilon} = \frac{N_{\mathrm{RREA}}}{7.3\ \mathrm{MeV}} \exp(-\varepsilon/7.3\ \mathrm{MeV}), \tag{9}$$

where 7.3 MeV is the mean energy of RREA EEDF obtained from MC results by *Dwyer et al.* [2012]. This equation also shows that we should expect the energy distribution to follow the exponential cut-off $\exp(-\epsilon/7.3\,MeV)$. In Appendix A, we give a complete derivation of the energy spectrum using the mean energy found from GEANT4 simulations.

In Appendix B we express the X-ray fluence distribution $f_\gamma(\epsilon_\gamma)$ (equation (B1)) by the bremsstrahlung differential cross section $\frac{d\sigma_\gamma}{d\epsilon_\gamma}(\epsilon_{re}, \epsilon_\gamma)$. We then show that the ratio of photons to electrons $N_\gamma(\epsilon_{th})/N_{re}(\epsilon_{th})$ can be expressed analytically by the ratio of the respective fluence distributions or by the bremsstrahlung coefficient $\alpha(\epsilon_{th})$ and avalanche time $\tau(E)$, where the differential bremsstrahlung cross section integrated from the lower energy threshold of integration of the electrons $\epsilon_{th}$ to infinity. In order to find the amount of electrons required to produce a given flux of photons, we can determine $\alpha(\epsilon_{th})$ and $\tau(E)$ empirically from simulation results.

### 1.3. Feedback

As previously explained, MC modeling has been used to explain the observations by the RHESSI and BATSE satellites. Results have indicated that between $10^{14}$ and $10^{17}$, runaway electrons are required to produce a TGF, assuming a production altitude of 15 km and that the ratio of bremsstrahlung photons to electrons is roughly 1. According to *Carlson et al.* [2008], we can assume that cosmic rays produce a maximum seed population of $10^6$ energetic electrons. Furthermore, we can expect an electric potential of 100 MV to be available in a large thundercloud, which would roughly correspond to 100 MV/7.3 MV = 13.9 avalanche lengths or a maximum multiplication $e^{13.9} \approx 10^6$ runaway electrons per seed electron. Combining this we get a multiplication of $10^{12}$, which is 5 orders of magnitude lower than the required number of electrons from an average RHESSI TGF produced at 15 km altitude. In addition, *Dwyer* [2008] made calculations on the initiation of RREAs from extensive air showers (EAS) and steady state background radiation, both mainly a product of cosmic rays. He found that neither of them is very likely to explain TGFs by its own. Thus, the high number of electrons required to produce a TGF cannot be explained by RREA multiplication alone. In response to this, two leading theories have been presented. *Dwyer* [2003] suggested that the feedback mechanism could provide further multiplication and thus explain the production of TGFs. Another possible solution has been presented by *Celestin and Pasko* [2011]. They show that seed electrons with energies on the order of 60 keV can be produced in the vicinity of the tips of lightning leaders by streamers and be further accelerated in the potential drops in front of lightning leader tips. They found that this process was capable of producing $10^{17}$ energetic electrons. In this paper, we will only examine the feedback mechanism as modeled by GEANT4.

During the initial avalanche, electrons traveling upward in the opposite direction of the electric field will produce many energetic bremsstrahlung photons. Some of these photons will either Compton backscatter or produce pairs of electrons and positrons. If the backscattered photons produce additional runaway electrons, through Compton scattering or photoelectric absorption, inside the strong electric field, they can initiate secondary avalanches. If pair production occurs and positron-electron pairs are produced, the positrons will quickly accelerate downward along the electric field, in the opposite direction of the electrons. If the positrons travel without annihilating, they may also initiate secondary avalanches through electron by positron elastic scattering (Bhabha scattering).

Due to the alignment of the electric field and the low probability of particle interaction, only positrons or photons can backscatter and initiate secondary avalanches. The two mechanisms responsible for feedback are called X-ray feedback and positron feedback, depending on the backscattered particle. In addition, secondary effects such as the products of positron annihilation or bremsstrahlung photons emitted from backscattering positrons, also have an effect [see *Dwyer*, 2007] but will not be distinguished from the primary feedback processes in this paper.

In *Dwyer* [2003], feedback multiplication was quantified by the feedback factor $\gamma$. The feedback factor describes the rate at which RREAs are multiplied. The relation is given as a common ratio in a geometric series and is derived in Appendix C. For $\gamma < 1$ and a number of avalanches $n \to \infty$, the total number of electrons converges to

$$N_n = N_{re}/(1-\gamma), \tag{10}$$

for $\gamma = 1$,

$$N_n = N_{re}n, \tag{11}$$

and for $\gamma > 1$ and $n >> 1$, the sum can be expressed as

$$N_n = N_{re}\gamma^n. \tag{12}$$

The feedback time $\tau_{fb}$ is the average time for runaway electrons and the backward propagating positrons or photons to complete one round trip within the avalanche region [*Dwyer*, 2012]. The total number of electrons produced $N_{tot}$ is then, for $\gamma > 1$, given as

$$N_{tot} = N_{RREA}\gamma^{t/\tau_{fb}} = N_0\gamma^{(t/\tau_{fb})}e^{(t/\tau(E))}, \tag{13}$$

where $N_0$ is the number of seed electrons, $t$ is the time, and $\tau(E)$ is still the avalanche time [*Dwyer*, 2003]. If $\gamma$ rises above 1, the electron multiplication process becomes unstable and the number of avalanches will increase exponentially.

## 2. The Monte Carlo Model

MC modeling has been widely used to test and constrain theoretical models. The RREA process has been studied in great detail [*Gurevich et al.*, 1992; *Lehtinen et al.*, 1999; *Babich et al.*, 2001; *Coleman and Dwyer*, 2006; *Celestin and Pasko*, 2010], the electron multiplication in streamer tip electric fields has also been studied by multiple models [*Celestin and Pasko*, 2010, 2011; *Chanrion and Neubert*, 2010]. However, studies that discuss the feedback process have solely been obtained from the model by *Dwyer* [2003].

The GEANT4 programming toolkit supplies a detailed library of physics processes concerning the interaction of particles with matter and is widely used in particle physics as well as studies in medical and space science [*Geant4 collaboration*, 2012b]. As GEANT4 is a well-established toolkit used for particle interactions, we suggest that it is an ideal candidate to study particle interactions in the atmosphere. Several studies concerning TGF production and propagation through the atmosphere have been compared to GEANT4. For example, *Carlson et al.* [2007] presented a new set of constraints on TGF production and *Østgaard et al.* [2008] used GEANT4 as a reference for comparison of bremsstrahlung emissions. In addition, *Smith et al.* [2010] used GEANT3 for reference of atmospheric absorption of TGF propagation. In this study, we will use the well-established GEANT4 toolkit to study the RREA and the relativistic feedback process.

The RREA process was first presented by *Gurevich et al.* [1992] and has since been studied in great detail. We will use results from these studies as benchmark to examine the accuracy of our simulations. Then, we will use GEANT4 to study the feedback process and the feedback factor to test and validate the results presented by *Dwyer* [2003, 2007, 2012], which to our knowledge has not been validated by independent studies before.

### 2.1. Physics Lists

GEANT4 has a very wide range of applications covering extremely energetic particle physics in the PeV range to low-energy physics in the hundreds of eV range. Depending on the energy regime in which simulations are performed, GEANT4 provides several models in the form of physics lists, which includes the physics processes that are required (see the physics reference manual for a detailed description [*Geant4 collaboration*, 2012a]). The RREA and the feedback process take place in the 10 keV to 100 MeV energy range. In this energy regime, several models have been validated. This study will compare two physics lists: (1) The standard model (chapter 8 in the Physics Reference Manual) or “the Low- and High-Energy Parameterization” (LHEP) list, which is developed by the Electromagnetic Standard Physics Working Group used for 1 keV to 10 PeV interactions; (2) The Livermore physics model (chapter 9 in the Physics Reference Manual) or “the Low Background Experiment” (LBE) list, which is developed by the Low Energy Electromagnetic Physics Working Group and used for 250 eV to 100 GeV (bremsstrahlung process included down to 10 eV) [*Geant4 collaboration*, 2013a, 2013b, 2012a]. In both lists, all important particle interactions that contribute to ionization of the atmosphere have been included. It also includes pair production, bremsstrahlung, elastic MÅller, and Bhabha scattering with free electrons and Compton scattering.

It should be noted that the previous studies that use GEANT4 have not discussed the use of physics lists. An important result of the present study is that the choice of physics lists has significant effects on the modeling results.

### 2.2. Simulation Setup

We have modified an MC code developed by the GEANT4 collaboration to model the electron avalanches in air under the influence of a homogeneous electric field. The following geometric and atmospheric composition standards have been used; A cylindrical volume of height $L = 10$ km and a diameter $d = 3000$ km, air consisting of 78.08% nitrogen, 20.95% oxygen, and 0.97% argon at standard sea level pressure and density, $2.684 \cdot 10^{25}$ molecules $/\mathrm{m}^3$. The electric field strength is chosen between 300 and 2500 kV/m, which is $\approx$0.1–0.75 $E_{cb}$. The electric field is extended vertically from a distance $z = 200$ m above the lower boundary of the cylinder to a distance determined from the amount of electron multiplication that occurs. We use the results by *Coleman and Dwyer* [2006] to make an assumption on the rate of multiplication and choose the vertical extension of the electric field to be between $3\lambda$ and $10\lambda$ according to equation (6). Unless otherwise stated we initiate each simulation with a continuous monoenergetic beam of 500 seed electrons of 1 MeV in the antiparallel direction of the electric field.

### 2.3. Selection Criteria

To study the RREA process we choose to include all particles that have a momentum along the initial trajectory of the avalanche (forward) $p > 0.0$. This also includes secondary particles produced by forward propagating photons and positrons. Each electron is then tracked and sampled with both time and location. We sample the electrons as they pass through 10 equally spaced screens inside and at the end of the electric field region. In an electric field extending from $z = 200$ m to $z = 400$ m, these screens are positioned at $z$ =(220.0, 240.0, 260.0, 280.0, 300.0, 320.0, 340.0, 360.0, 380.0, and 400.0) m. We also sample the electrons within each $8 \cdot 10^{-11}$ s interval from start to stop of the simulation. The time step of the electrons is accurate to $1 \cdot 10^{-12}$ s.

The feedback mechanism was studied using a different selection method. We tag every electron with a number corresponding to the generation each electron belongs to. The primary generation tag (1) is given to all electrons that pass through the final screen of the electric field region and are identified to be a part of the initial avalanche (see the previous paragraph). As opposed to the selection of RREAs, when studying feedback we store the position, momentum, and kinetic energy of the electrons that are produced from backscattering photons and positrons. These electrons are then supplied as seed electrons, with the stored data as initial conditions, in the next simulation and their secondary particles will in turn be given the second generation tag (2), and so on. With this method we must run one simulation per generation of feedback we wish to study.

## 3. Results

### 3.1. RREA Results

The avalanche length must be calculated using only runaway electrons. We use two methods to determine the energy range of the electrons to be included. For the time-dependent selection, we sample all electrons within each $8 \cdot 10^{-11}$ s time interval, from start to stop of the simulation. To only include runaway electrons, we need to determine the lower energy threshold for electrons to be accelerated for each electric field strength. This energy threshold can be determined from the average stopping power (see Figure 1) that is opposed to the electric field. However, electrons are rarely in perfect alignment with the electric field and, thus, the scattering of the electrons must be taken into account. The computation of the actual runaway threshold is fully described by *Lehtinen et al.* [1999]. For an electric field strength of $\sim$436 kV/m, the runaway threshold is $\varepsilon_{th} = 549$ keV, and for an electric field strength of $\sim$3270 kV/m the runaway threshold is $\varepsilon_{th} = 24$ keV [see *Lehtinen et al.*, 1999, Table 2].

For location-dependent simulations, we use a different assumption. Due to collisions and interactions between the electrons and the atmosphere, the electrons with low energy and some in the intermediate energy range (from a few hundred keV to a few MeV) will lose energy and stop before they pass through the screens. The runaway electrons, however, will by definition continue to accelerate and pass through the screens.

The electron multiplication was studied using the standard simulation setup. Simulations were initiated by a monoenergetic beam of 500 seed electrons of 1 MeV. The electrons were then tracked continuously, both in $8 \cdot 10^{-11}$ s time intervals and as they pass through 10 equally spaced screens in the electric field region.

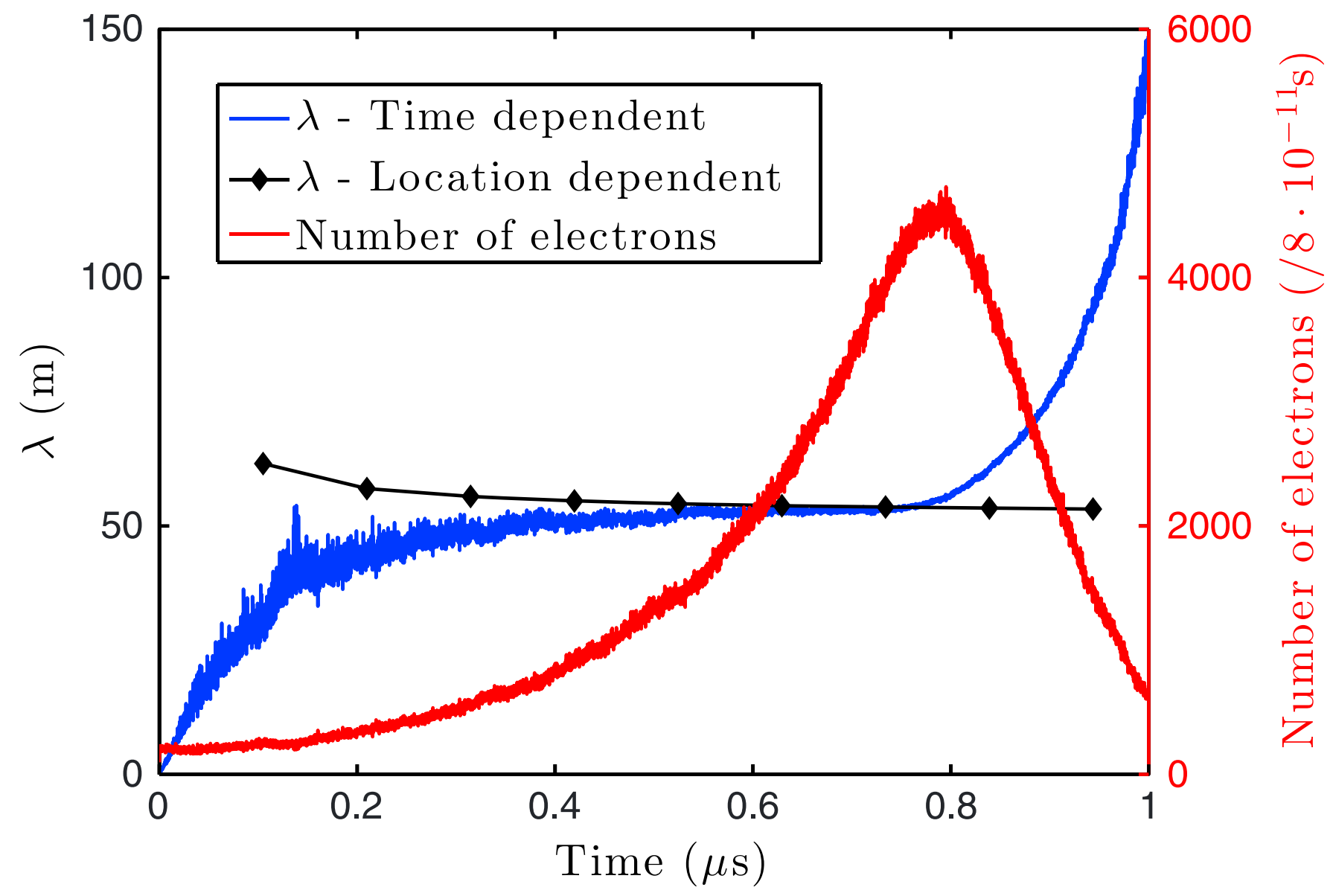

**Figure 2.** Electron multiplication with time (red line) and the corresponding avalanche lengths $\lambda$ by time dependency (blue line) and location dependency (black line). The correlation between time and distance is given by the average speed of the developing avalanche of 0.89$c$ *Coleman and Dwyer* [2006]. This case was simulated using the LBE physics list.

By rearranging equation (4) we can find the avalanche length as a function of the number of runaway electrons and their location, $N(z)$, or by time intervals, $N(t)$,

$$\lambda(z) = \frac{z}{\ln\left(N(z)/N_0\right)}, \qquad \tau(t) = \frac{t}{\ln\left(N(t)/N_0\right)} \tag{14}$$

where $z$ and $t$ are the distance and time and $N_0$ is the number of seed electrons. In Figure 2, we show the exponential increase of runaway electrons with time (red line) and the corresponding avalanche lengths from time-dependant (blue line) and location-dependant (black line) results. Although the selection is done in both time and position, we calculate the avalanche length by distance. For time-dependant simulations the avalanche length is determined by the average position of the electrons in the direction of the electric field. After a few avalanche lengths the RREAs reach a state of steady multiplication were $\lambda(z)$ does not vary with increasing time or distance.

We ran the simulations at intervals of electric field strength of 100 kV/m, between 300 kV/m and 2500 kV/m and then estimated the avalanche length $\lambda(E)$ with respect to the strength of the electric field. The results are shown in Figure 3 as red triangles for the LBE simulations and blue triangles for the LHEP results. The avalanche length obtained from LBE and the LHEP results, valid for 310 kV/m $< E <$ 2500 kV/m, can be fitted respectively by

$$\lambda(E)_{\mathrm{LBE}} = \frac{7400\ \mathrm{kV}}{E - 298.0\ \mathrm{kV/m}} \quad \text{and} \quad \lambda(E)_{\mathrm{LHEP}} = \frac{9770\ \mathrm{kV}}{E - 285\ \mathrm{kV/m}}, \tag{15}$$

where $E$ is the electric field strength. These functions are shown in Figure 3 (red and blue, respectively) as well as the function presented by *Coleman and Dwyer* [2006], (equation (6) (black line)).

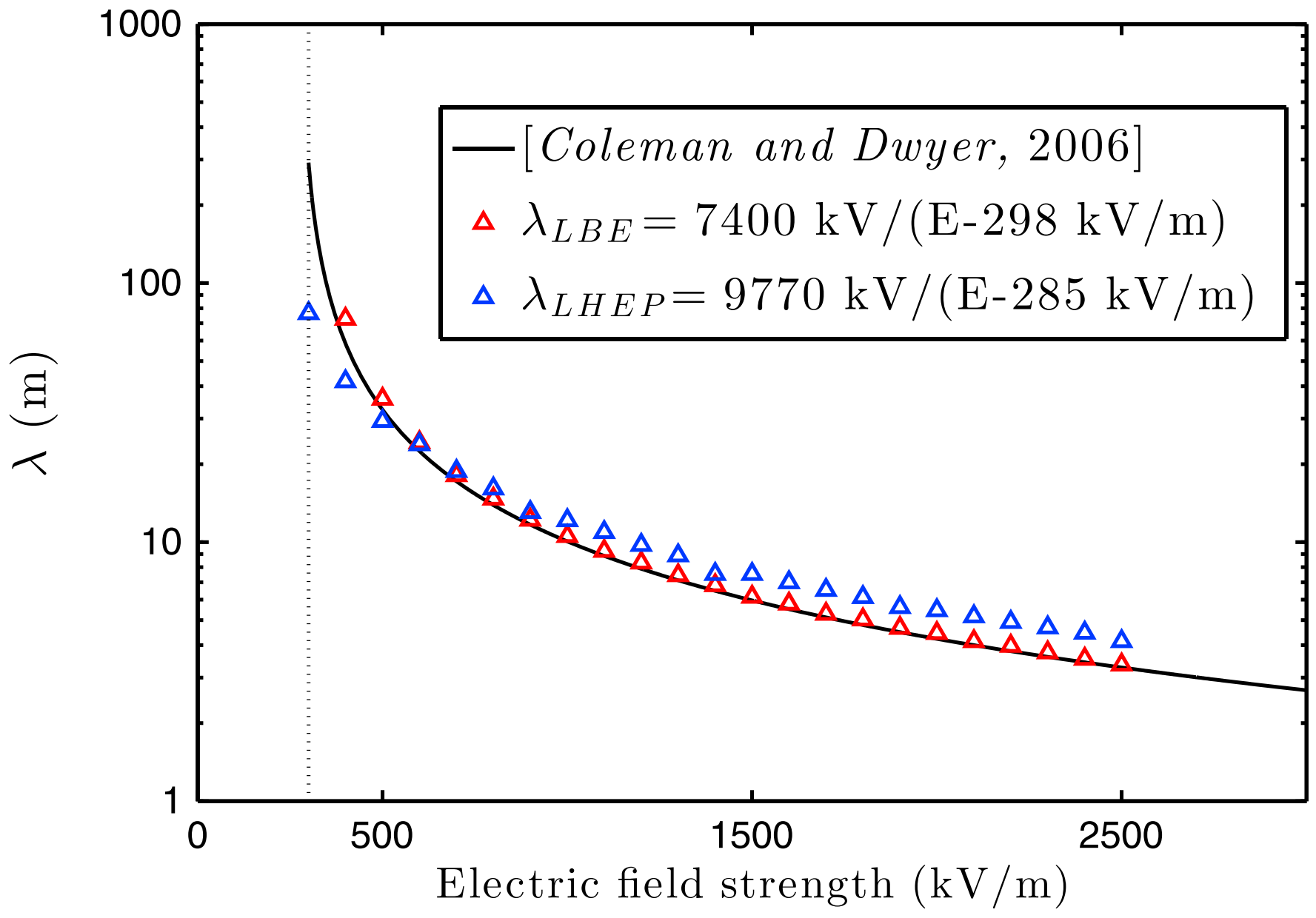

**Figure 3.** The avalanche length $\lambda$ with respect to electric field strength. The LBE and LHEP results are marked as red and blue triangles, respectively. The continuous black line follows equation (6), which is the avalanche length determined by *Coleman and Dwyer* [2006]. In addition, the dotted black line marks the minimum electric field threshold $E_{\mathrm{th}}$ = 284 kV/m.

Comparing the results obtained from GEANT4 simulations we see that the avalanche lengths from the LBE results agree to within ±5% with the LHEP results at electric fields between 500 and 700 kV/m. At electric fields below 500 kV/m the LBE results tend to give much longer avalanche lengths than the LHEP results, with a maximum of approximately 60%. For stronger electric fields (> 600 kV/m) the difference is on average approximately 25% and the LBE results give shorter avalanche lengths than the LHEP results. By comparing the GEANT4 results to the results by *Coleman and Dwyer* [2006] (equation (6)), we also see that the LBE results which is on average only 5% above the results by Coleman and Dwyer, much closer than the avalanche length obtained from LHEP simulations, which differ by ±35%. This shows that the explicit production of low-energy electrons and photons used by the LBE physics list provide results closer to the estimates by *Coleman and Dwyer* [2006] than implementing the continuous energy loss functions as by the LHEP physics list.

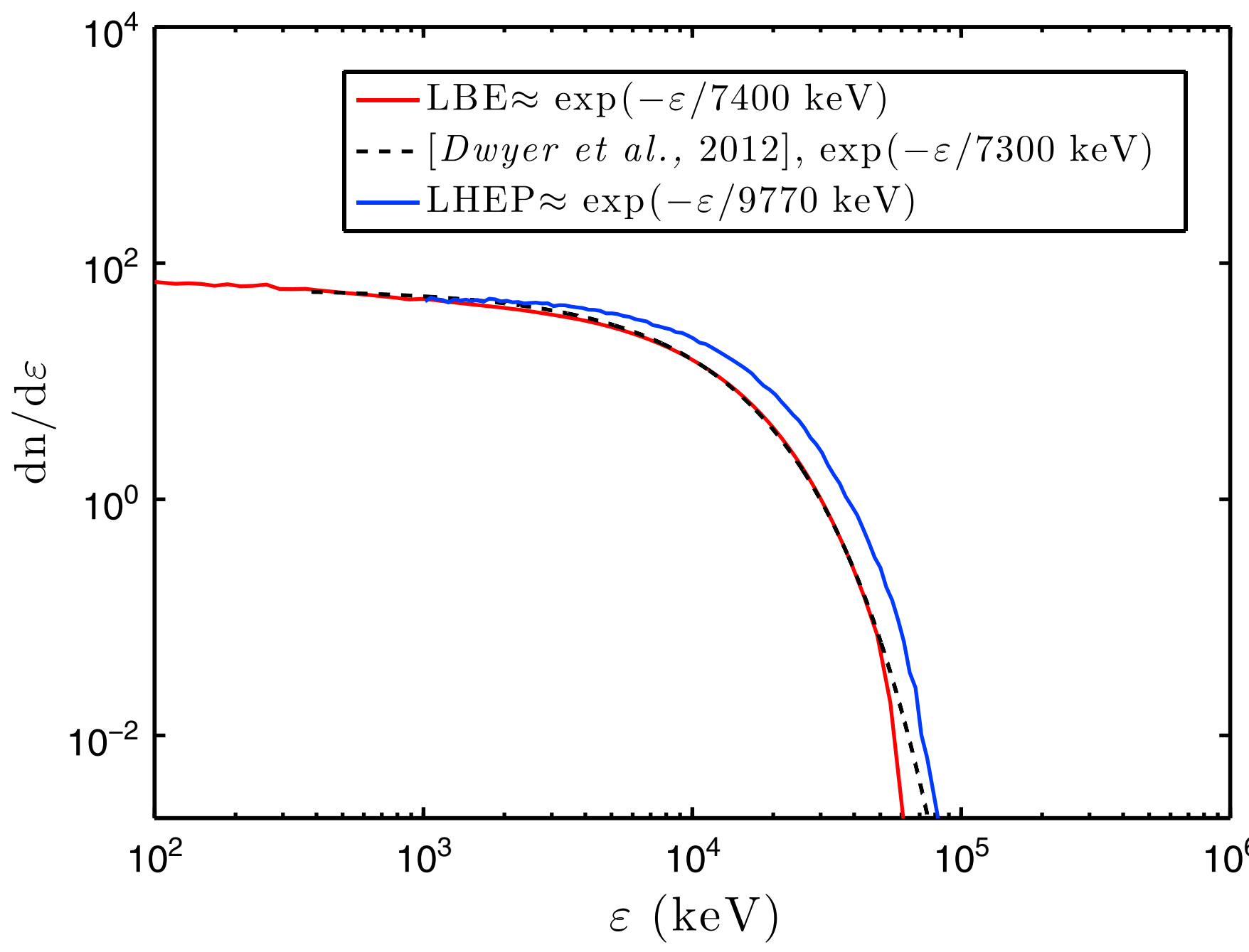

**Figure 4.** The electron energy distribution from simulations using the LBE physics list (red) and the LHEP physics list (blue). In addition, these results are compared to the exponential function exp($-\varepsilon/7300$) found by *Dwyer et al.* [2012] (dashed black). Note that there is a deviation at very high energies where GEANT4 results do not follow an exponential cutoff.

We also determined the avalanche time, $\tau(E)$, from LBE results,

$$\tau(E) = \frac{27.4 \text{ kV µs/m}}{E - 298 \text{ kV/m}} = \frac{\lambda(E)}{0.9c}, \quad (16)$$

which indicates that the average speed of the avalanche is ≈ 0.9*c* and is close to the estimate by *Coleman and Dwyer* [2006] of 0.89*c*. This also corresponds with the average distance propagated by the time-dependant simulations. The electron energy distribution function (EEDF) of the runaway electrons was also studied using both the LBE and the LHEP physics lists. Electric fields were chosen at intervals of 50 kV/m between 350 and 500 kV/m and at intervals of 100 kV/m between 500 and 2500 kV/m. The EEDF was calculated using 400,000 electrons at each interval.

Several studies have indicated that the EEDF above a few hundred keV can be described by an exponential cutoff function [*Lehtinen et al.*, 1999; *Celestin and Pasko*, 2010; *Dwyer and Babich*, 2011]. In *Dwyer and Babich* [2011] this cutoff was determined to be best fit by the exponential function exp($-\varepsilon/7300$ keV). In Figure 4, we show the results obtained from the LBE (red line) and LHEP (blue line) physics lists. In addition, the analytical 7300 keV cutoff is indicated as the dashed blue line. The LBE distribution corresponds well with previous estimates and follows an ≈ 7400 keV cutoff. However, the LHEP results are again not in agreement and show a much harder cutoff at ≈ 9770 keV. Note that in both cases the energy spectrum at very high energies fall off quicker than the exponential function. This is due to data processor limitations, which prevent the distribution to reach a steady state at very high energies due to very large number of particles in the simulations. When the electric field strength is close to the electric field threshold, ≈ 300 kV/m < E < 350 kV/m, the distributions have a much softer spectrum. As for the avalanche lengths, the mean energy of the electron distribution is dependent on the energy range of the runaway electrons in the distribution. The two methods gave substantially different results with mean energy of the time-dependant method being approximately 10% lower than the location-dependant method.

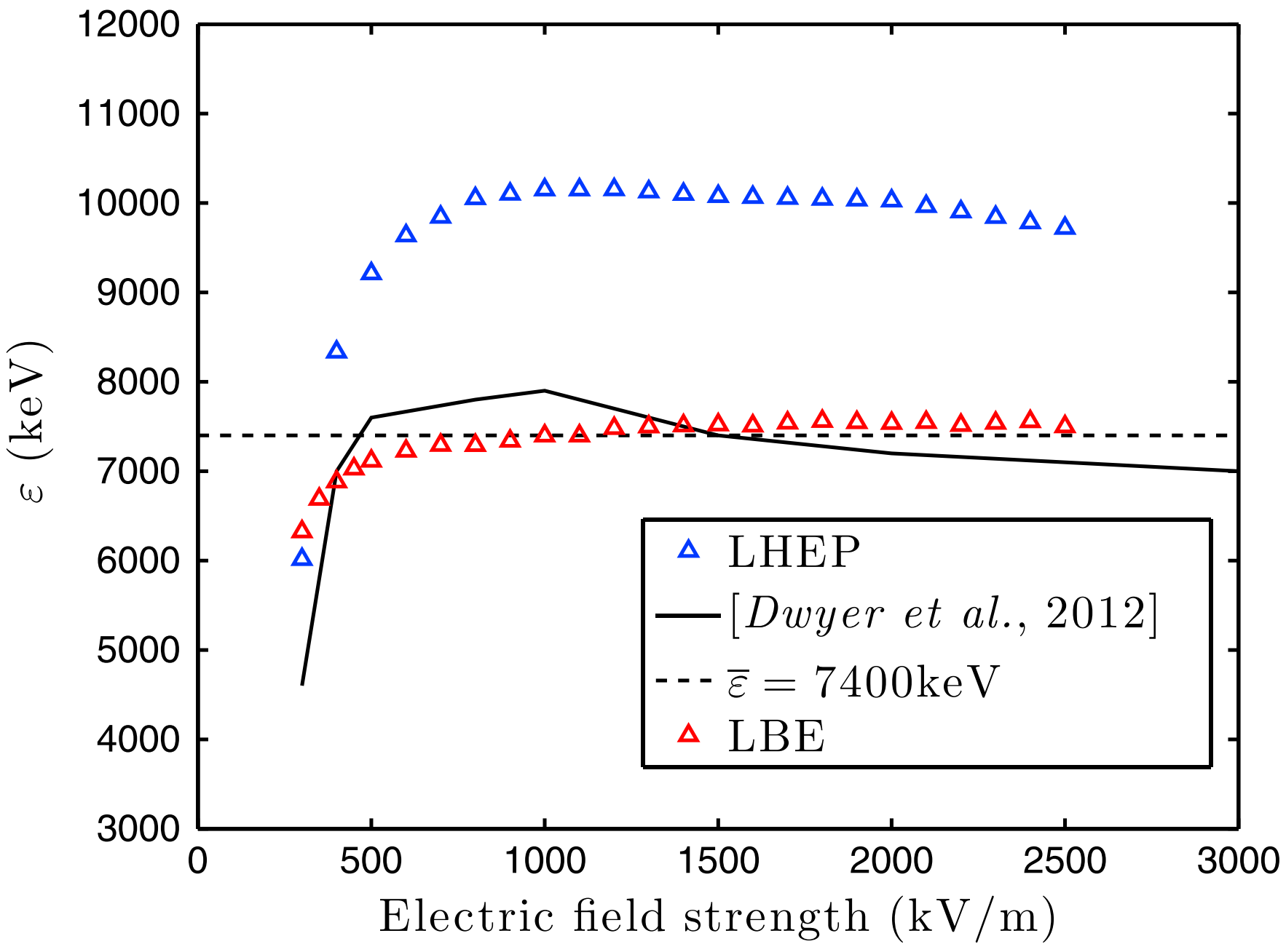

**Figure 5.** The mean energy of runaway electrons in an RREA with respect to electric field strength. The red and blue triangles again indicate the LBE results and the LHEP results, respectively. They are compared to the continuous black line, which represents the mean energies found by Dwyer [*Dwyer et al.*, 2012, Figure 3]. The dashed line represents, the average mean energy, $\bar{\varepsilon} = 7400$ keV found from LBE results.

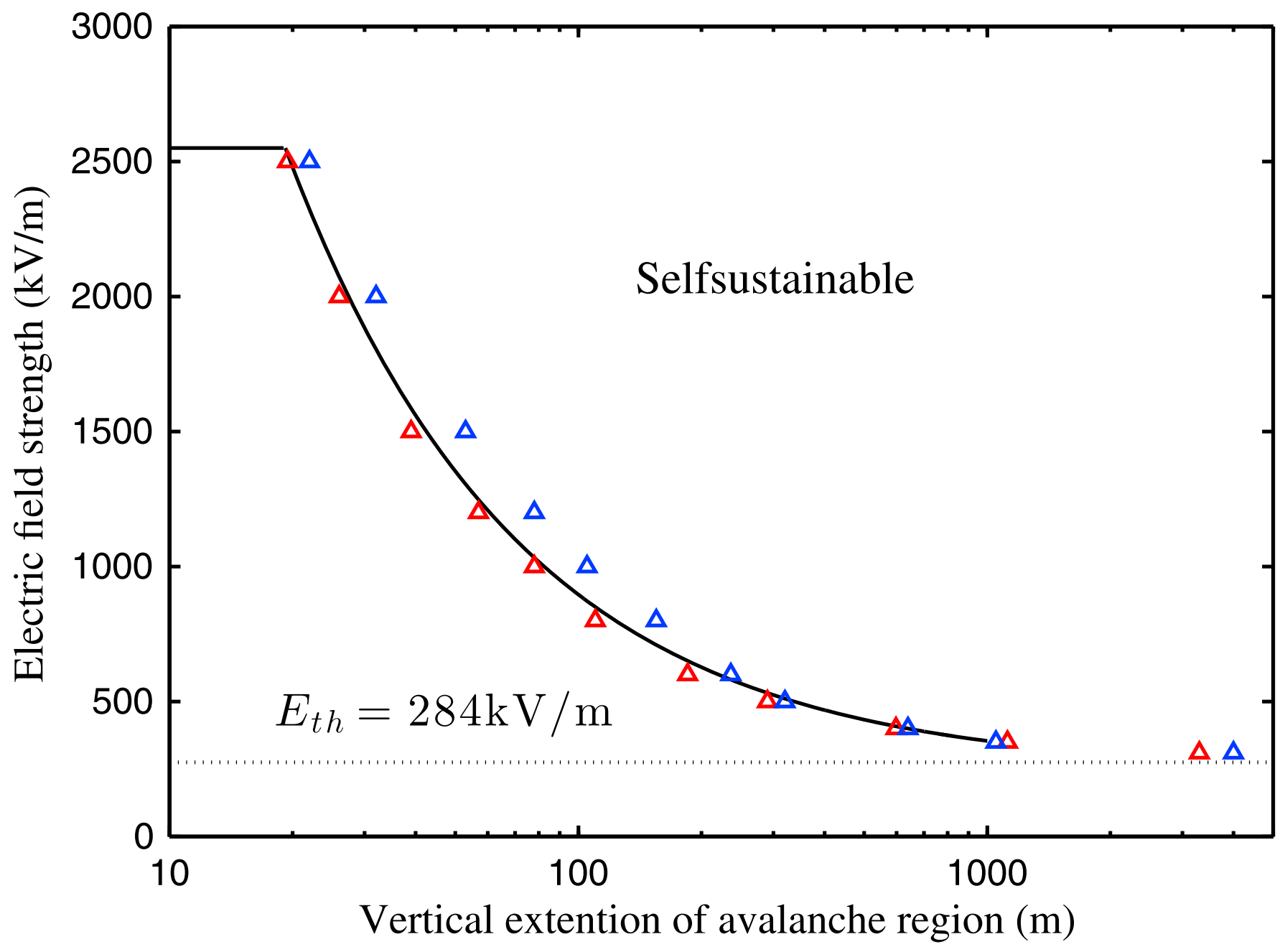

**Figure 6.** This figure shows the electric field strength versus the vertical extension of the avalanche region. At a given electric field strength, the vertical extent of the avalanche region has an upper limit where the multiplication process becomes self sustainable, and a complete or partial discharge will quickly occur. This limit was determined in *Dwyer* [2003] to be when the feedback factor $\gamma$ is equal to 1 and is indicated by the continuous black line. The results obtained from LBE and LHEP simulations are indicated by the red and blue triangles, respectively.

We choose to implement the location-dependent selection method as it is similar to the selection method used by *Dwyer* [2004].

The mean energy of the RREA distributions with respect to the electric field strength can be seen in Figure 5. We compare the LBE and the LHEP results to the results presented in *Dwyer and Babich* [2011]. The average energy resulting from LBE simulations, 7400 keV, is in good agreement with the 7300 keV from Dwyers results. This mean energy also corresponds to the best fit to the EEDF and to the avalanche length given by equation (15). The mean energy of LHEP simulations is approximately 9700 keV, which is 33% larger than previous estimates. A difference is also seen for electric fields above $\approx$1200 kV/m, where LBE results are stable close to 7500 keV and LHEP results show a weak decrease from $\approx$10, 200 to 9650 keV. This decrease of mean energy at stronger electric fields is also seen in the results of *Dwyer et al.* [2012].

### 3.2. Feedback

The feedback factor quantifies the increase or decrease of RREA by the relation between the number of REs in an initial RREA and the sum of REs produced by secondary RREAs. The secondary RREAs are electron avalanches that have been initiated from backscattering photons and positrons from the initial RREA. This relation is derived in Appendix C. In our simulations we store the position, momentum, and kinetic energy of the electrons produced by backscattering photons and positrons from the initial RREA. These data are then used to initiate the secondary avalanches in a separate simulation and again store the third generation seed electrons. For every simulation we also track all electrons passing through a screen at the end of the avalanche region, thus finding the relation between each RREA and its secondaries. We then determine the feedback factor $\gamma$ or the rate of feedback, using equations (10)–(12).

The feedback mechanism becomes unstable and increases the rate of RREA multiplication exponentially when the feedback factor rises above 1 [*Dwyer*, 2003]. In Figure 6, we show the threshold for the feedback multiplication to become unstable depending on the strength of the electric field and the vertical extension of the avalanche region at sea level density and pressure. The red and blue triangles show the results from the LBE and the LHEP simulations, respectively. For electric fields close to the electric field threshold ($<$ 500 kV/m) the LBE results show that 18% shorter avalanche regions than that in the case of LHEP are required for feedback to become unstable. At 500 kV/m the results are comparable with a difference of $\pm$5%. However, for fields stronger than 500 kV/m the difference quickly rises to 25%, which is considerable. For very strong fields close to the maximum field for feedback to become unstable (2550 kV/m from *Dwyer* [2003], for sea level density and pressure), the difference becomes slightly less pronounced, but still close to 20% longer avalanche regions are required. If we compare these results to those presented in *Dwyer* [2003] (black curve) we see that the LBE results are in good agreement with Dwyers results.

### 3.3. Photon to Electron Ratio

To study the ratio of the produced bremsstrahlung photons to the energetic electrons in a RREA, we use only the LBE physics list. Every simulation is initiated with a monoenergetic beam of 500 seed electrons of 1 MeV. At each interval of electric field strength, 400, 800, 1200, 1600, and 2000 kV/m, we let the avalanche develop over both five and seven avalanche lengths to examine the distributions of electrons and photons at the end of the avalanche regions. We then use equation (B6), with different choices for the energy threshold of integration ($\varepsilon_{\text{th}}$) for both electrons and photons to determine $N_\gamma/N_e$. Figure 7 shows the ratio $N_\gamma/N_e$ as a function of the energy threshold of integration. The darker colors signify development over $5\lambda$ and the brighter colors over $7\lambda$. A high-energy electron is less likely to transfer a large portion of its energy, through the bremsstrahlung process, to the produced photon. This corresponds to the drop in the photon to electron ratio with higher-energy thresholds. In addition, the ratio decreases with stronger electric fields because the electron multiplication expressed by the avalanche time $\tau(E)$ has a $1/E$ dependency. Another important feature is that the ratio is independent on the lifetime of the avalanche once steady state of the EEDF is reached. This is seen as the darker ($5\lambda$) points near perfectly overlap the brighter ($7\lambda$) points. However, this is not seen for very high energy thresholds of integration and this can be due to the relatively low number of electrons and photons in this energy range.

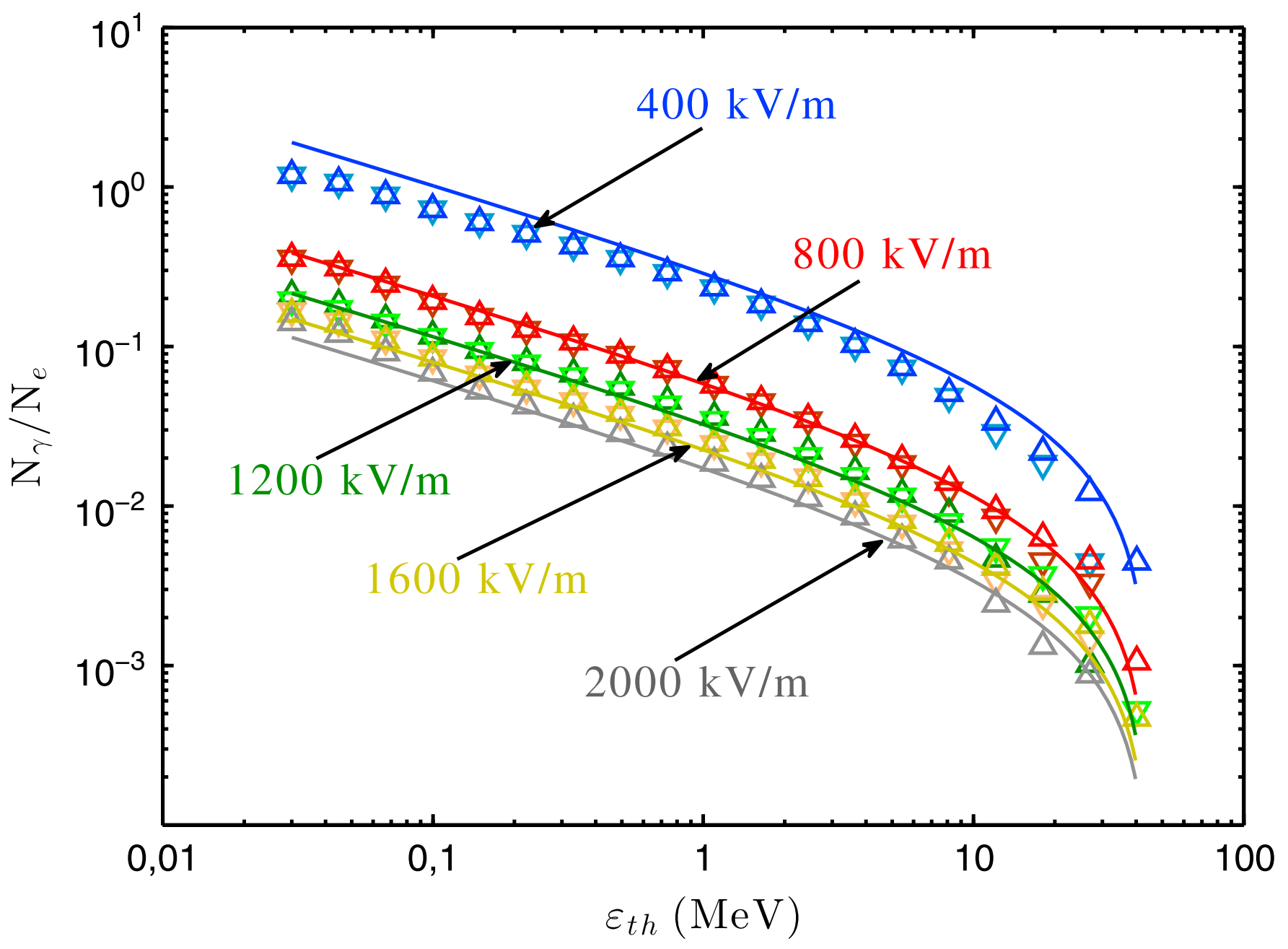

**Figure 7.** The ratio of bremsstrahlung photons to runaway electrons $N_\gamma/N_e$ from GEANT4 simulations. The points show the ratio obtained for given electric field strengths, and the brightness indicates the simulated distance of the avalanche, where $5\lambda$ is bright and $7\lambda$ is dark. The curves follow the analytical expression $N_\gamma/N_e = \alpha(\varepsilon_{\text{th}}) \cdot \tau(E)$, where $\alpha(\varepsilon_{\text{th}})$ is given in equation (17) and $\tau(E)$ is given in equation (16).

By rearranging equation (8) the bremsstrahlung production coefficient $\alpha$ can be expressed in terms of the ratio $N_\gamma/N_e$ and the avalanche time $\tau(E)$ (see Appendix B for a full derivation). As the ratio $N_\gamma/N_e$ and $\tau(E)$ are equally dependent on the electric field, $\alpha$ loses this dependency and is only dependent on the energy threshold $\varepsilon_{\text{th}}$. To determine $\alpha$ we then multiply the simulation results for $N_\gamma/N_e$ by the avalanche time $\tau(E)$ for the respective electric fields. The bremsstrahlung coefficient $\alpha(\varepsilon_{\text{th}})$, with respect to the energy thresholds of integration $\varepsilon_{\text{th}}$, valid for 0.1 MeV $< \varepsilon_{\text{th}} <$ 60.0 MeV, is found empirically to follow:

$$\alpha(\varepsilon_{\text{th}}) = \frac{1.258\ \left[\text{MeV}^{1/2}\ \mu\text{s}^{-1}\right]}{\sqrt{\varepsilon_{\text{th}}}} - 0.1874\ [\mu\text{s}^{-1}] \qquad (17)$$

where $\varepsilon_{\text{th}}$ is the energy threshold of integration in equation (B6) given in MeV and $\alpha(\varepsilon_{\text{th}})$ is given in $\mu\text{s}^{-1}$. The results are shown in Figure 8 where the colored triangles show the ratio $N_\gamma/N_e$ obtained from simulation results multiplied by the avalanche length $\tau(E)$ and the solid black curve is $\alpha(\varepsilon_{\text{th}})$ from equation (17). To confirm the result we also plot the ratio expressed by $\alpha(\varepsilon_{\text{th}}) \cdot \tau(E)$ on top of the simulation results in Figure 7 and find that the results are in agreement.

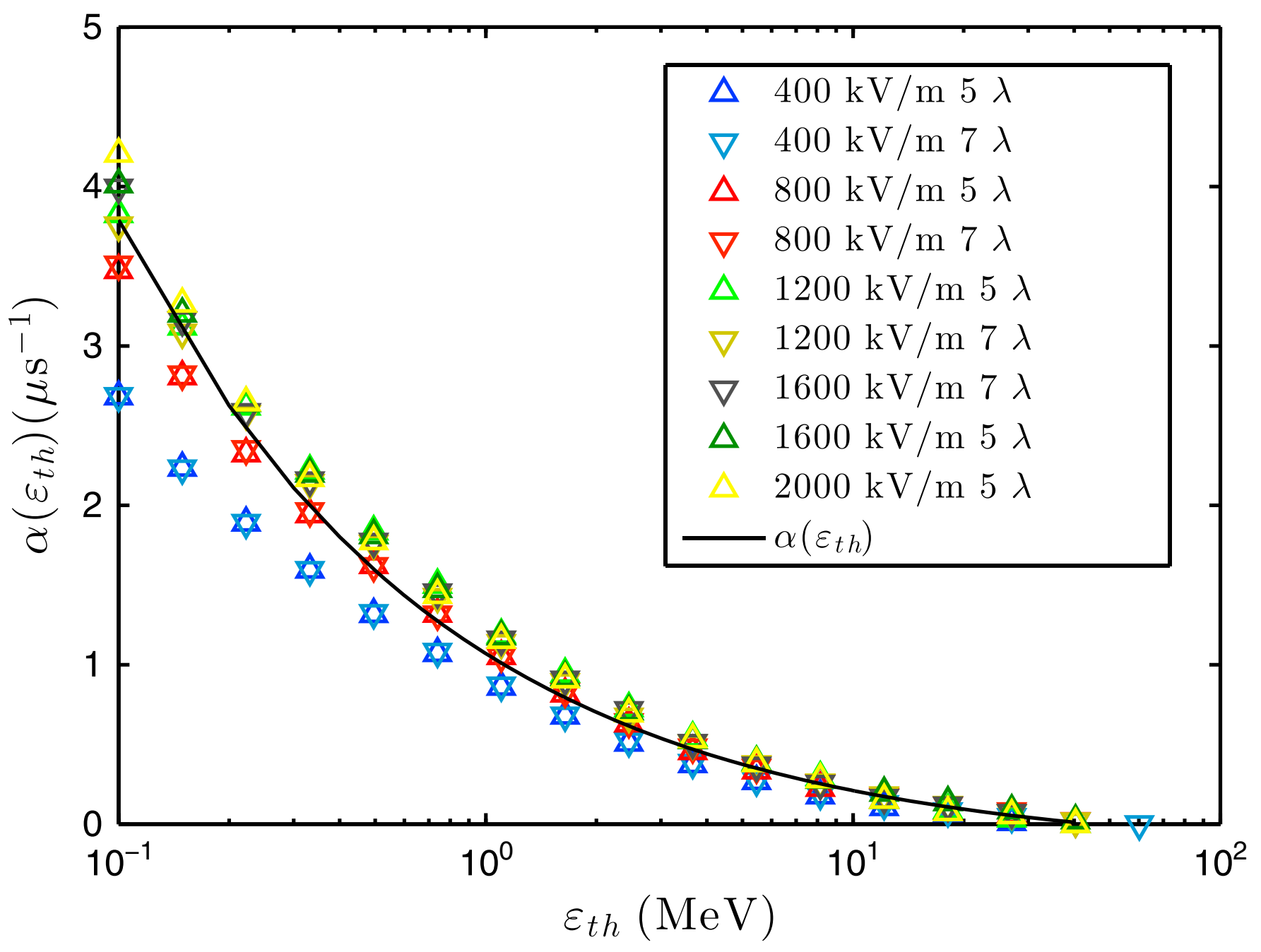

**Figure 8.** The bremsstrahlung coefficient $\alpha(\varepsilon_{\text{th}}) = (N_\gamma/N_e)/\tau(E)$ as a function of the energy threshold of integration $\varepsilon_{\text{th}}$. The colored triangles indicate the results from simulations for given energy boundaries, but for different electric field strengths and the continuous black line is the best fit function (equation (17)). In addition, for each electric field strength the brighter color indicates a simulated distance of $5\lambda$ and the darker color $7\lambda$.

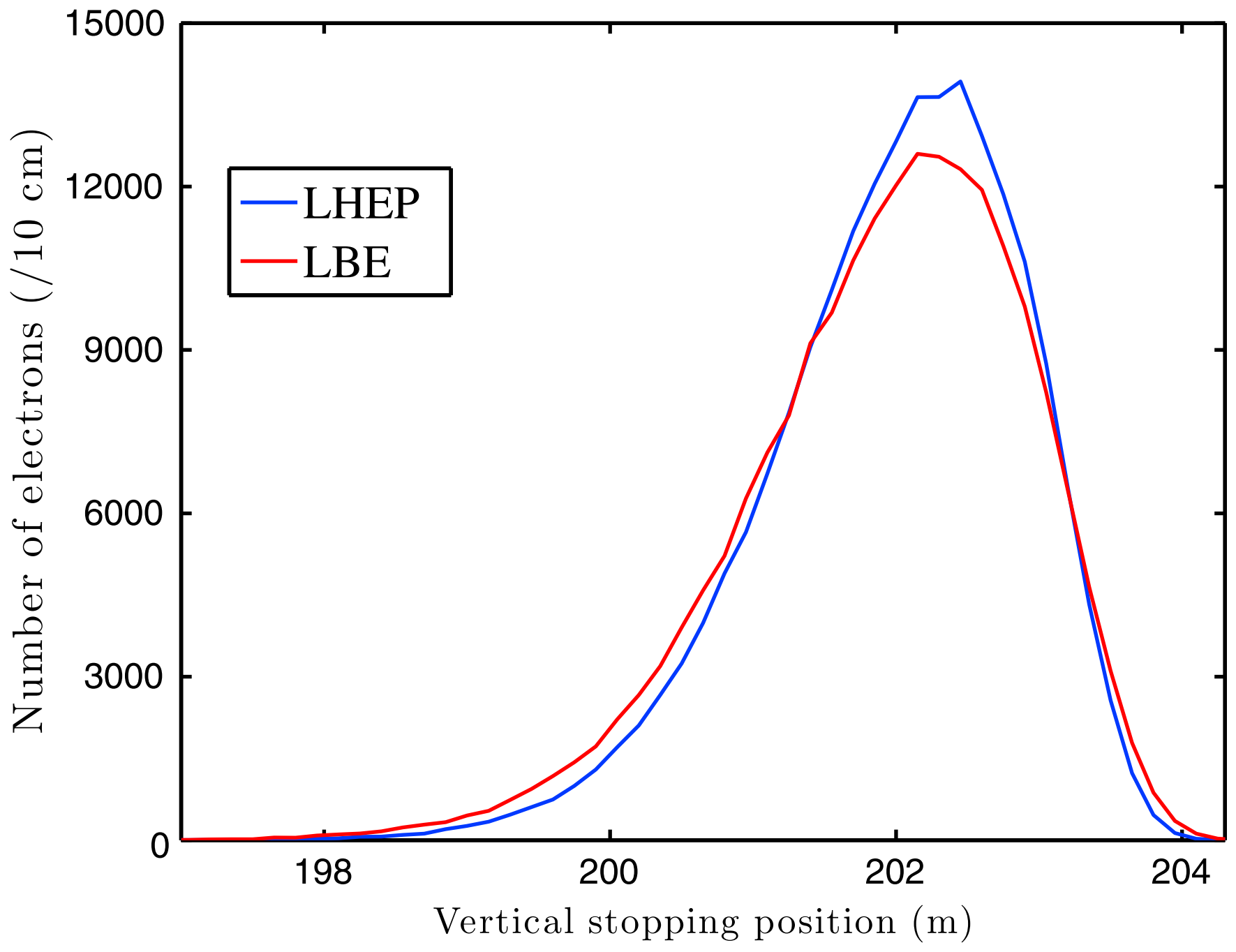

**Figure 9.** The stopping position of 50,000 electrons of 1 MeV per 10 cm without the influence of an electric field. The LBE results are indicated by the red color, and the LHEP results are indicated by the blue color. Note that the initial vertical position of the electrons is at 200 m.

For a lower energy threshold of integration equal to 1 MeV, the ratio $N_\gamma/N_e$ is found to be 0.23 for an electric field strength of $E = 400$ kV/m. For stronger electric fields, between 800 and 2000 kV/m, we find that the ratio is roughly in the range 0.1–0.01. This indicate that if TGFs are produced in very strong electric fields, as a result of the feedback mechanism, the required number of electrons at the source is 1–2 order of magnitude higher than the number of photons. However, if TGFs are produced in weaker electric fields close to the lower electric field threshold (284 kV/m), the ratio of bremsstrahlung photons to electrons become closer to 1.

## 4. Discussion

### 4.1. Physics List Comparison

An important result obtained from the GEANT4 MC simulations is the significant difference found by using the LBE or LHEP physics list. Although both lists have been validated in the energy range we have studied, the results are substantially different. The main difference between these two physics lists is the implementation of the continuous and discrete energy losses of electrons and positrons due to ionization and bremsstrahlung. When using the LHEP list, the energy loss function is introduced for energies below 1 keV. However, while using the LBE list, the energy loss function is introduced below 250 eV for ionization and below 10 eV for bremsstrahlung. Above these thresholds the energy loss is simulated explicitly through the production of photons, electrons, and positrons. In addition, the cross sections in the LBE physics list make direct use of shell cross-section data [*Geant4 collaboration*, 2012a]. The cross sections in the LHEP physics list does not take the shell cross-section data into account. It is clear from these differences that the LBE physics list contains more accurate descriptions of low-energy interactions, in particular, between 250 eV and 1 keV.

To determine the effect of theses differences on the energy loss of the electrons, we initiated a continuous monoenergetic beam of 50,000 electrons of 1 MeV at a vertical position of 200 m without the influence of an electric field. We then found the vertical stopping position of each individual electron. The result is shown in Figure 9. On average, the electrons in the case of the LHEP simulations make it further than in the case of the LBE simulations. The mean stopping position of electrons is 10 cm (4%) further in the LHEP case as compared with the LBE case. We can infer from this that less energy is lost on average as a result of the continuous loss functions by the LHEP simulations as compared to the LBE simulations.

In section 3.1, our results show that the avalanche length on average is 25% longer for strong electric fields and 60% longer close to the electric field threshold. In addition, as less energy is lost through interactions with low energy particles when using the LHEP list, the mean energy of the RREA electron distribution becomes correspondingly larger (see Figure 5). The mean difference was shown to be $\approx$ 32% larger.

From the comparisons between our results and the previous results from independent MC models, we can conclude that it is likely that the LBE physics list is more accurate when studying RREAs and feedback. In fact, results obtained from simulations using the standard LHEP model greatly underestimate the energy loss and electron multiplication of RREAs.

### 4.2. RREA

The avalanche length or rate of runaway electron multiplication, the mean energy, and the electron energy distribution function (EEDF) are key features necessary to discuss in order to validate GEANT4 as a tool to simulate RREAs. In section 3.1 we discuss the use of a location- or time-dependent selection method and find that the simulation results are different depending on which method we implement. In order to determine the extent of the difference between the two methods, we compare the corresponding normalized spectra obtained. This can be seen in Figure 10. At higher energies the time-dependent (red) and location-dependent (blue) results match perfectly. At lower energies, however, the location-dependent selection method includes only the fraction that managed to run away and pass through the screen at which position they are sampled. The time-dependent selection method acts as a camera, taking a snapshot of the system that then includes all low-energy electrons that are free at that particular moment.

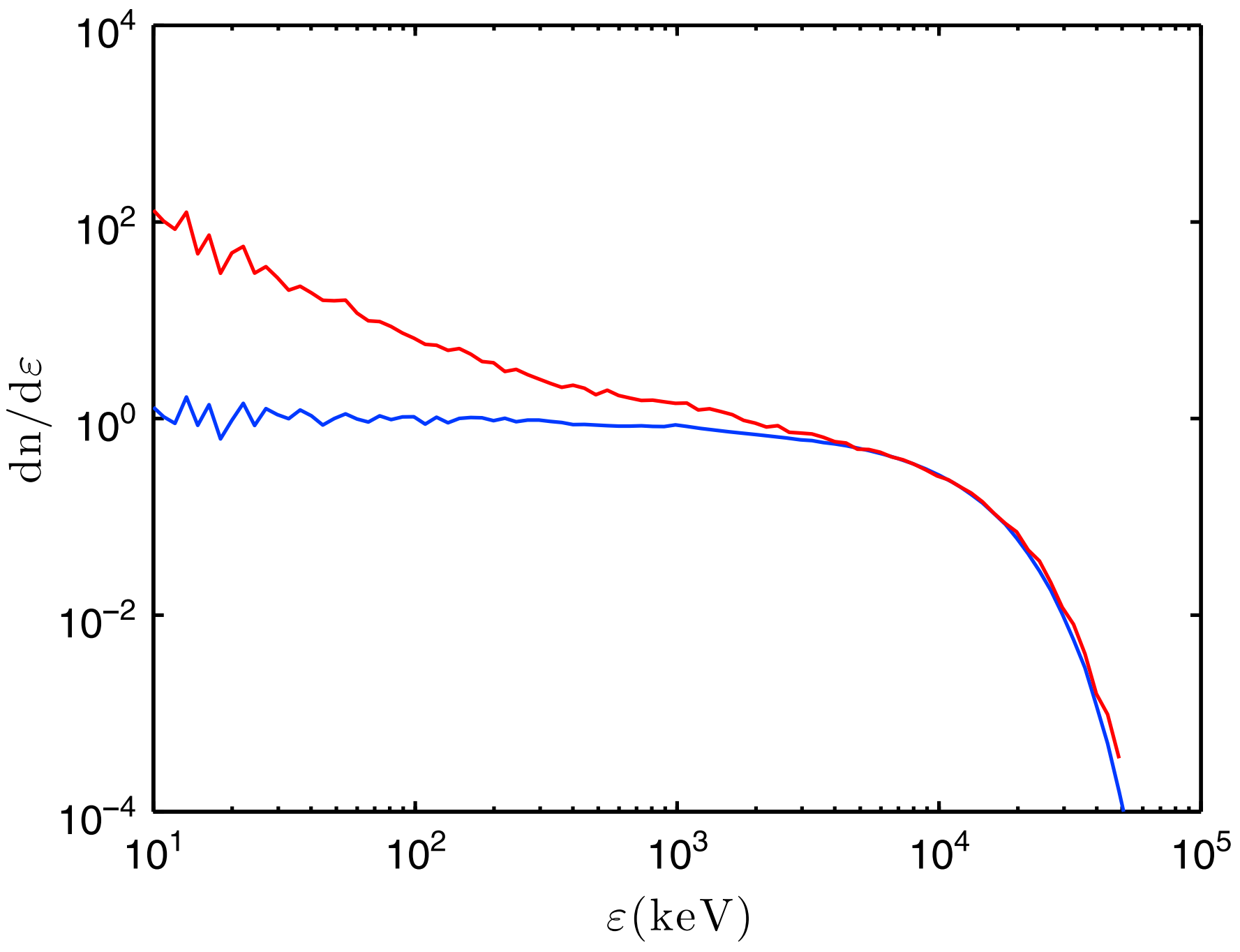

**Figure 10.** The energy distribution of RREAs simulated in an electric field of 400 kV/m. The red graph indicates the time-dependent simulation results, and the blue graph indicates the distance-dependent results.

For the purpose of studying the EEDF and the mean energy of RREAs we choose to define the runaway electrons as all electrons that are accelerated by the electric field. By this definition the location-dependent method directly samples all runaway electrons, as all the REs are accelerated through the final screen of the avalanche region. The time-dependent method, however, includes many low-energy electrons and some intermediate-energy electrons that will eventually stop due to interactions and collisions, which will result in a 10% lower mean energy even if the runaway energy threshold is taken into account.

*Coleman and Dwyer* [2006] found that the exponential growth of runaway electrons in a RREA could be well described by equation (6). This is also supported by a comparison with independent results from several authors [*Dwyer et al.*, 2012]. The results obtained from GEANT4 modeling, using the LBE physics list, are also in very good agreement with previous studies. The difference between our results and equation (6) is less than 5% for weak electric field strengths ($<$ 600 kV/m) and less than 3% for stronger electric fields (600–2500 kV/m).

In order to study the RREAs in more detail we also compared the mean energy and the EEDFs to existing results. The mean energy of electrons in a RREA can be expressed as the net energy gained over one avalanche length by the electric field and lost by the friction force of the atmosphere. In a homogeneous electric field at sea level pressure and density, the mean energy and the energy distribution of RREAs reach a steady state after a few avalanche lengths. The total mean energy in this steady state, averaging over all electric field strengths between 300 and 2500 kV/m, was found to be $\bar{\varepsilon}_{\text{LBE}} \simeq$ 7400 keV. This is in very good agreement with previous estimates of $\bar{\varepsilon} =$ 7300 keV by *Dwyer et al.* [2012]. However, a detailed comparison shows that for weak, 300–600 kV/m, and strong, 1300–2500 kV/m, electric fields, the mean energy is slightly higher than previous results. While in the intermediate range, 600–1300 kV/m, the mean energy is slightly lower. In other words the total mean energy is in good agreement, but shows less variations with the strength of the electric field.

From results presented in *Dwyer and Babich* [2011], we expected the EEDF to follow an exponential cutoff $\exp(-\varepsilon/\bar{\varepsilon})$, where $\varepsilon$ is the energy of the electrons and $\bar{\varepsilon}$ is the average mean energy of RREAs. As expressed in the previous paragraph we found the total mean energy to be $\bar{\varepsilon}_{\text{LBE}} \simeq$ 7400 keV. Using the location-dependent method of simulation, we calculated the cutoff above 500 keV and found that the EEDF was best fit by the exponential function $\exp(-\varepsilon/7440\text{ keV})$, which is close to the expected value. For energies above $\approx$ 50 MeV, the energy distribution falls off slightly faster and no longer follows the exponential cutoff. However, the limitations of data handling of extremely large number of particles make the number density of electrons with energies in the several tens of MeV range low. This may influence the spectrum and make it difficult to obtain a distribution that is in steady state at very high energies.

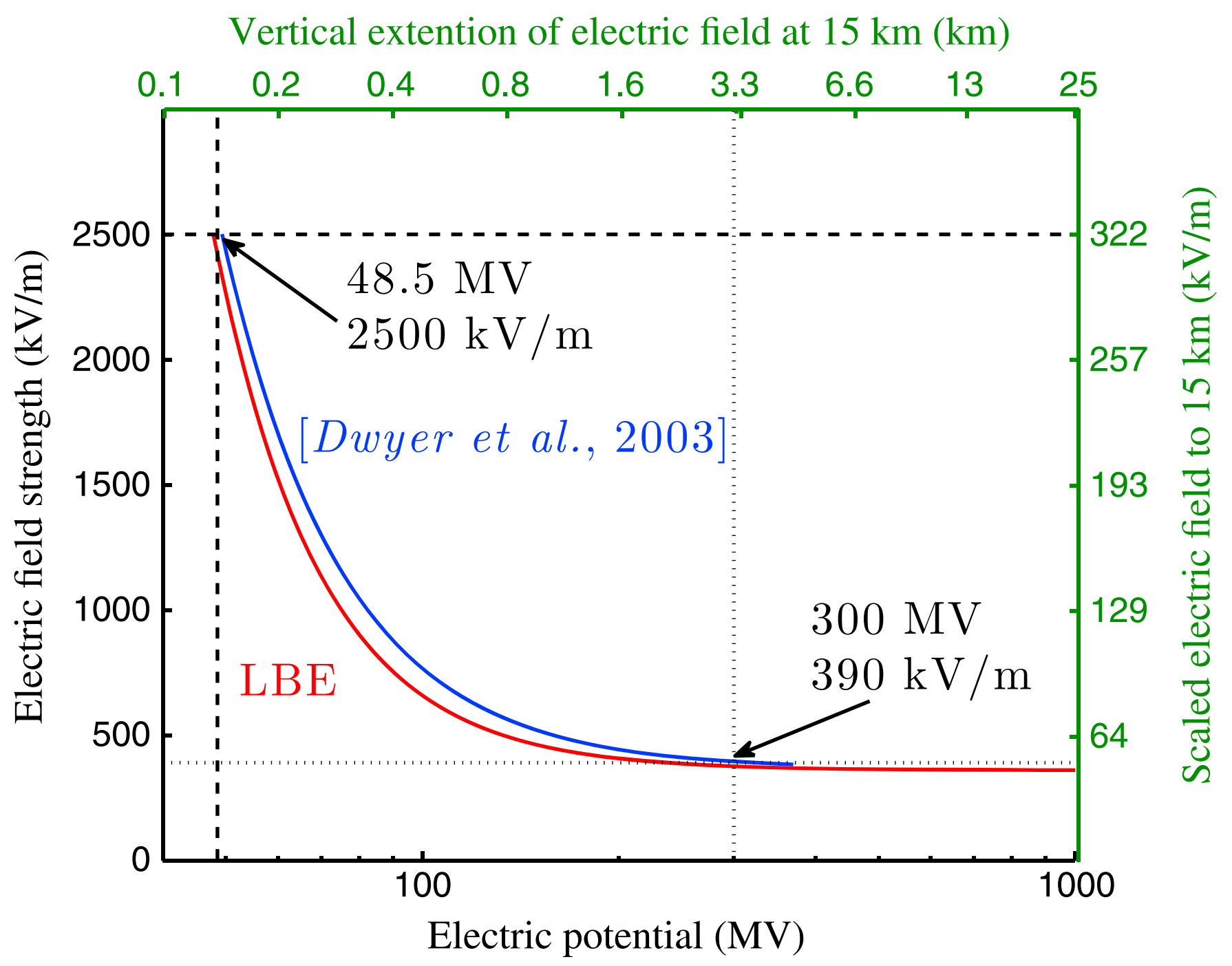

**Figure 11.** The feedback threshold expressed by the electric field strength and the corresponding electric potential. The LBE results (red) are in good agreement with the estimates from *Dwyer* [2003] (blue). The minimum potential required for feedback to become self sustainable is determined to be at 48.5 MV and at 2500 kV/m (dashed line). Very large thunderstorms are believed to produce electric potential differences on the order of 300 MV, which would require a minimum electric field strength of 390 kV/m (dotted line). The green *x* and *y* axes show the electric field strength and corresponding vertical extension scaled to 15 km altitude.

This shows that our model is well equipped to simulate RREAs in a homogeneous electric field with strength close to the conventional breakdown field, $E_{cb} = 3200$ kV/m. If the electric field strength becomes larger than the critical breakdown field, electrons with energies lower than the runaway threshold become important and the LBE physics list may then become insufficient. These results are therefore important in terms of validating GEANT4 MC simulations as an important tool to study RREAs. However, as we have shown, the implementation of the physics lists is crucial in order to obtain accurate results.

### 4.3. Feedback

An important subject of this paper is to test the effect of the feedback process on electron multiplication during RREA development in Earth's atmosphere. To do this we wish to find the conditions required for the feedback process to become self sustainable ($\gamma > 1$). The primary conditions to push the feedback factor, $\gamma$, above 1 were defined in *Dwyer* [2003] to be the strength of the electric field and its vertical extension. As shown in Figure 6, the results obtained with LBE simulations are in very good agreement with the conditions presented by *Dwyer* [2003], in particular, for weak electric fields close to the electric field threshold.

At the production altitude of TGFs ($\approx$ 15 km), the electric field must be scaled by the atmospheric density. However, the electric potential required to accelerate and multiply the REs remains constant. Figure 11 shows the feedback threshold expressed by the electric field strength and the corresponding electric potential. In addition, we show the electric field strength and the corresponding vertical extension scaled to 15 km altitude on the green *x* and *y* axes, respectively. *Dwyer* [2003] estimated that this potential must be on the order of 50 MV and increasing with weaker electric fields to several hundred Mega Volts. GEANT4 simulations confirm these results. We have estimated that an electric potential of 48.5 MV (19.4 m) at an electric field strength of 2500 kV/m to $\approx$ 300 MV ($\sim$770 m) at 390 kV/m is required.

Typical thunderstorms have electric potential differences of up to 100 MV [*Carlson et al.*, 2009]; however, it has been proposed that larger storms may produce potential differences in the order of 300 MV. Balloon soundings through thundercloud systems measured maximum potential differences in IC regions to be in the range of 40–130 MV [*Marshall and Stolzenburg*, 2001]. This is above the feedback potential thresholds found for electric field strengths of 400–2500 kV/m (270–48.5 MV). However, for the feedback mechanism to be self sustainable, these potential differences must be limited to local regions of the thundercloud in order to sustain the relatively strong electric fields. Marshall and Stolzenburg concluded that, although they measured maximum electric fields of only one third of the conventional breakdown threshold, such strong electric fields appear in relatively small regions and have very short lifetimes.

From GEANT4 simulations we can confirm the results presented by Dwyer and also conclude that it cannot be ruled out that the feedback mechanism play a role in the production of TGFs given the right conditions.

### 4.4. Photon to Electron Ratio

The ratio of bremsstrahlung photons to electrons is important to determine the amount of electrons required to produce a TGF. In Figure 8, we show that $\alpha(\varepsilon_{\rm th})$ depends weakly on the strength of the electric field and can be expressed as a function of the energy threshold of integration. Furthermore, we show (see also Appendix B) that the ratio of bremsstrahlung photons to runaway electrons for a given electric field strength, can be expressed by

$$\frac{N_\gamma}{N_e} = \alpha(\varepsilon_{\rm th})\tau(E), \tag{18}$$

where $\tau(E)$ and $\alpha(\varepsilon_{\rm th})$ are given in equations (16) and (17) and $\varepsilon_{\rm th}$ is the energy threshold of integration of electrons and photons.

This threshold can be determined by relating the force exerted on the electrons by the electric field to the friction force experienced by electrons with a given energy (Figure 1). The intersection between the two forces will give the average minimum energy required for an electron to be runaway. Using this energy as the energy threshold of integration will give an approximation of the expected photon to electron ratio. For an electric field of 400 kV/m, where the energy threshold of integration is estimated to be $\approx$ 549 keV by *Lehtinen et al.* [1999] and *Celestin and Pasko* [2010], the photon to electron ratio is roughly 0.8. When the electric fields become stronger, although the energy threshold of integration becomes lower, the ratio becomes closer to 0.1. These results are in good agreement with the general assumption of 1 to 0.1 photon to electron source ratio.

## 5. Conclusion

1. GEANT4 is widely used as a toolkit to validate modeling results. However, the use of different physics lists has not been discussed in previous studies concerning the production of TGFs. We have shown that the choice of physics list is crucial to obtain correct results.
2. We have obtained the first detailed results concerning the RREA process using GEANT4. The LBE physics list (Livermore model) provides results that are in very good agreement with previous studies. As results concerning RREAs are well established, our results are important to validate GEANT4 as a toolkit to study electron multiplication in the Earth's atmosphere.
3. This paper also presents the first independent study of the feedback mechanism presented in *Dwyer* [2003]. Our results confirm the results presented by *Dwyer* [2003] and constrain the conditions under which the feedback mechanism may play a role in the production of high-energy particles in thunderstorms.
4. The ratio of bremsstrahlung photons to runaway electrons $N_\gamma/N_{\rm re}$ in electric fields between 400 and 2000 kV/m was found to be between 1 and 0.1. This can be calculated using the analytical expression presented in equation (18), where the bremsstrahlung coefficient $\alpha(\varepsilon_{\rm th})$ has been determined empirically from GEANT4 simulations and is given in equation (17).

## Appendix A: The Energy Distribution of RREAs

To find the energy distribution of the RREA electrons, we start by restating the avalanche length of a RREA in an electric field of strength between 310 kV/m < $E$ < 2500 kV/m from equation (15)

$$\lambda = \frac{7400\ \text{keV}}{eE - F_d} \tag{A1}$$

where 7400 keV is the mean energy of the runaway electrons in a RREA determined from our Monte Carlo simulations. *Gurevich et al.* [1992] showed that the number of runaway electrons in one avalanche increases with time $t$ and distance $z$ and can be expressed by distance as

$$\frac{\mathrm{d}N}{\mathrm{d}z} = \frac{1}{\lambda}N, \tag{A2}$$

where $N$ is the number of electrons at a given distance $z$ from the start of the avalanche. We then integrate equation (A2) from the number of seed electrons at the start of an avalanche $N_0$ to the total number of electrons $N_{\rm RREA}$ at the end of the avalanche region $L$ and get

$$\int_{N_0}^{N_{\rm RREA}} \frac{\mathrm{d}N}{N} = \int_0^L \frac{\mathrm{d}z}{\lambda}. \tag{A3}$$

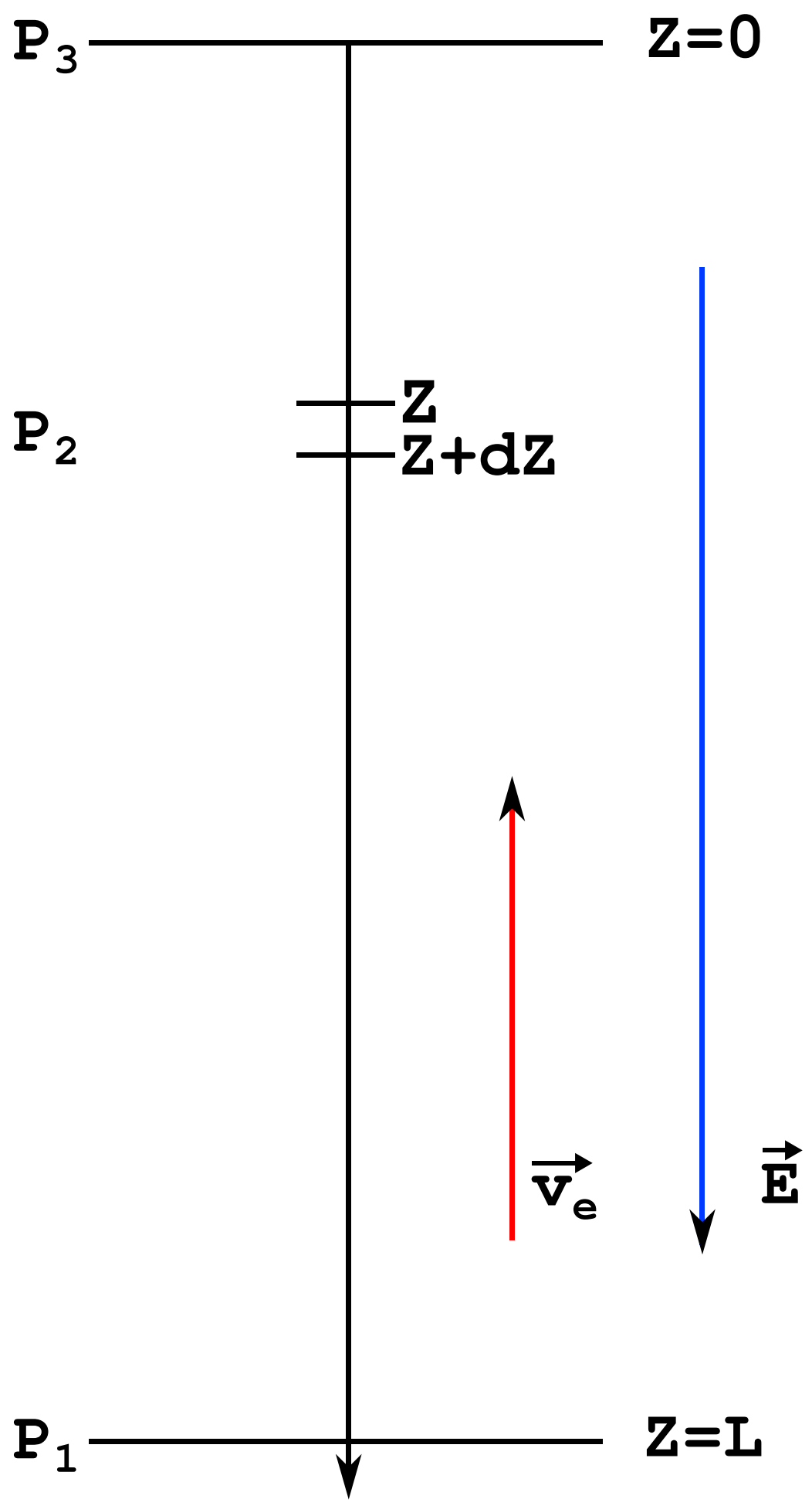

**Figure A1.** A schematic representation of a system where the electrons move with a velocity $\vec{v}_e$ in an electric field $\vec{E}$ and thus creating an avalanche in the region from $Z = 0$ to $Z = L$. Three points of interest are also marked in the figure as points $P_1$, $P_2$, and $P_3$.

The total number of particles produced in a RREA is then given by

$$N_{\mathrm{RREA}} = N_o \exp\left(\int_0^L \frac{\mathrm{d}z}{\lambda}\right) = N_o \exp\left(\frac{L}{\lambda}\right). \tag{A4}$$

Figure A1 shows a schematic representation of the system used to derive the energy distribution. If we have a given number of seeding electrons $N_0$ at position $P_1$ at the start of the avalanche region $Z = L$ (Note that we have a positive direction along the electric field, and thus the avalanche develops in the negative direction). The number of runaway electrons at point $P_2$ at a distance $Z$ from the start of the avalanche is then

$$N(Z) = N_0 \exp\left(\frac{L-Z}{\lambda}\right). \tag{A5}$$

The number of electrons moving a distance d$Z$ from $Z + \mathrm{d}Z$ to $Z$ is given by deriving equation (A5):

$$-\frac{\mathrm{d}N(Z)}{\mathrm{d}Z} = -\frac{N_0}{\lambda} \exp\left(\frac{L-Z}{\lambda}\right). \tag{A6}$$

The negative sign is due to moving in the negative direction from $Z + \mathrm{d}Z$ to $Z$. The change in the number of electrons with respect to distance is then

$$\mathrm{d}N(Z) = N_0 \exp\left(\frac{L-Z}{\lambda}\right)\frac{\mathrm{d}Z}{\lambda} \tag{A7}$$

The average kinetic energy gained by an electron moving a distance $Z$ in an electric field $E$ is given by

$$\varepsilon = Z(eE - F_d) \tag{A8}$$

where $F_d$ is the friction force experienced by the electron and is $\approx$ 218 keV/m. Rearranging equation (A8) gives us

$$Z = \frac{\varepsilon}{eE - F_d}. \tag{A9}$$

We then assume that the avalanche length can be described as

$$\lambda = \frac{7400\ \mathrm{keV}}{eE - F_d}, \tag{A10}$$

where the average electron energy 7400 keV is obtained empirically from LBE simulations. Dividing equation (A9) with equation (A10)

$$\frac{Z}{\lambda} = \frac{\varepsilon}{7400\ \mathrm{keV}} \tag{A11}$$

and then by derivation we see that

$$\frac{\mathrm{d}Z}{\lambda} = \frac{\mathrm{d}\varepsilon}{7400\ \mathrm{keV}}. \tag{A12}$$

Inserting equations (A11) and (A12) in equation (A7), we change dependence from distance $Z$ to electron energy $\varepsilon$

$$\mathrm{d}N(\varepsilon) = N_0 \exp\left(\frac{L}{\lambda} - \frac{\varepsilon}{7400\ \mathrm{keV}}\right)\frac{\mathrm{d}\varepsilon}{7400\ \mathrm{keV}} \tag{A13}$$

or in terms of $N_{\mathrm{RREA}}$

$$\mathrm{d}N(\varepsilon) = \frac{N_{\mathrm{RREA}}}{7400\ \mathrm{keV}} \exp\left(-\frac{\varepsilon}{7400\ \mathrm{keV}}\right)\mathrm{d}\varepsilon. \tag{A14}$$

The energy distribution or the number of runaway electrons per unit energy is then

$$N_{\mathrm{re}} = \frac{\mathrm{d}N(\varepsilon)}{\mathrm{d}\varepsilon} = \frac{N_{\mathrm{RREA}}}{7400\ \mathrm{keV}} \exp\left(-\frac{\varepsilon}{7400\ \mathrm{keV}}\right). \tag{A15}$$

## Appendix B: The Bremsstrahlung Coefficient

The X-ray source spectrum $f_\gamma$ can be expressed by

$$f_\gamma(\varepsilon_\gamma) = N_n \int f_{\rm re}(\varepsilon_{\rm re}) \frac{d\sigma_\gamma}{d\varepsilon_\gamma}(\varepsilon_{re}, \varepsilon_\gamma) v(\varepsilon_{\rm re}) d\varepsilon_{\rm re}, \tag{B1}$$

where $N_n$ is the neutral gas density, $f_{\rm re}(\varepsilon_{\rm re})$ is the electron fluence distribution, $d\sigma_\gamma/d\varepsilon_\gamma$ is the differential bremsstrahlung cross section and $v(\varepsilon_{\rm re})$ is the velocity of electrons with energy $\varepsilon_{\rm re}$.

In Appendix A, the EEDF was shown analytically to follow an exponential cutoff. If the avalanche has developed sufficiently and is in steady state, the resulting X-ray spectrum $f_\gamma(\varepsilon_\gamma)$ does not depend on time. The ratio of the number photons to the number of runaway electrons then depend on the runaway electron multiplication as a function of electric field strength $E$ by $\tau(E)$, which is the avalanche time, and on the bremsstrahlung coefficient $\alpha$ integrated over all energies, and can be expressed by

$$\frac{N_\gamma(\varepsilon_{\rm th})}{N_{\rm re}(\varepsilon_{\rm th})} = \alpha(\varepsilon_{\rm th})\tau(E) = \frac{\int_{\varepsilon_{\rm th}}^{+\infty} f_\gamma(\varepsilon_\gamma) d\varepsilon_\gamma}{\int_{\varepsilon_{\rm th}}^{+\infty} f_{\rm re}(\varepsilon_{\rm re}) d\varepsilon_{\rm re}}, \tag{B2}$$

$$= \frac{N \int_{\varepsilon_{\rm th}}^{+\infty} \int_{\varepsilon_{\rm th}}^{+\infty} f_{\rm re}(\varepsilon_{\rm re}) \frac{d\sigma_\gamma}{d\varepsilon_\gamma}(\varepsilon_{\rm re}, \varepsilon_\gamma) v(\varepsilon_{\rm re}) d\varepsilon_{\rm re} d\varepsilon_\gamma}{\int_{\varepsilon_{\rm th}}^{+\infty} f_{\rm re}(\varepsilon_{\rm re}) d\varepsilon_{\rm re}}, \tag{B3}$$

$$= \frac{N \int_{\varepsilon_{\rm th}}^{+\infty} f_{\rm re}(\varepsilon_{\rm re}) v(\varepsilon_{\rm re}) \int_{\varepsilon_{\rm th}}^{+\infty} \frac{d\sigma_\gamma}{d\varepsilon_\gamma}(\varepsilon_{\rm re}, \varepsilon_\gamma) d\varepsilon_{\rm re} d\varepsilon_\gamma}{\int_{\varepsilon_{\rm th}}^{+\infty} f_{\rm re}(\varepsilon_{\rm re}) d\varepsilon_{\rm re}}, \tag{B4}$$

where $\varepsilon_{\rm th}$ is the lower energy threshold of integration of the electrons. We then let

$$\xi(\varepsilon_{\rm re}, \varepsilon_{\rm th}) = N_n v(\varepsilon_{\rm re}) \int_{\varepsilon_{\rm th}}^{+\infty} \frac{d\sigma_\gamma}{d\varepsilon_\gamma}(\varepsilon_{\rm re}, \varepsilon_\gamma) d\varepsilon_\gamma, \tag{B5}$$

and finally express the ratio of the number of photons to the number of electrons by

$$\frac{N_\gamma(\varepsilon_{\rm th})}{N_{\rm re}(\varepsilon_{\rm th})} = \alpha(\varepsilon_{\rm th})\tau(E) = \frac{\int_{\varepsilon_{\rm th}}^{+\infty} f_{\rm re}(\varepsilon_{\rm re}) \xi(\varepsilon_{\rm re}, \varepsilon_{\rm th}) d\varepsilon_{\rm re}}{\int_{\varepsilon_{\rm th}}^{+\infty} f_{\rm re}(\varepsilon_{\rm re}) d\varepsilon_{\rm re}}. \tag{B6}$$

Using equation (B6) and the avalanche multiplication $\tau(E)$ by equation (16), we can express the bremsstrahlung coefficient by

$$\alpha(\varepsilon_{\rm th}) = \frac{N_\gamma(\varepsilon_{\rm th})}{N_{\rm re}(\varepsilon_{\rm th})} \frac{1}{\tau(E)} = \frac{N_\gamma}{N_e} \frac{E - 298\,\text{kV/m}}{27.4\,\text{kV µs/m}}, \tag{B7}$$

where $E$ is the strength of the electric field.

## Appendix C: The Feedback Factor

If $N_n$ is the total number of electrons produced after $n$ number of secondary avalanche multiplications, the feedback factor $\gamma$ is given as the relation between the total number of electrons in the $n$th and $n$th − 1 avalanche

$$\gamma = \frac{N_n}{N_{n-1}}. \tag{C1}$$

We can then express the total number of electrons at the end of the avalanche as the sum of a geometric series using the number of electrons produced in the initial RREA $N_{\rm re}$ and the feedback factor $\gamma$

$$N_n = N_{\rm re} + N_{\rm re}\gamma + N_{\rm re}\gamma^2 + \ldots + N_{\rm re}\gamma^{n-1}, \qquad n = 1, 2, 3, \ldots \tag{C2}$$

$$\gamma N_n = N_{\rm re}\gamma + N_{\rm re}\gamma^2 + \ldots + N_{\rm re}\gamma^{n-1} + N_{\rm re}\gamma^n. \tag{C3}$$

We now get the sum of the $n$th partial sum by subtracting these two equations,

$$(1-\gamma)N_n = N_{\mathrm{re}}(1-\gamma^n), \tag{C4}$$

and for $\gamma \neq 1$ we get

$$N_n = \frac{N_{\mathrm{re}}(1-\gamma^n)}{1-\gamma}. \tag{C5}$$

If $\gamma < 1$ and $n \to \infty$, the total number of electrons converges to

$$N_n = \lim_{n\to\infty} \frac{N_{\mathrm{re}}(1-\gamma^n)}{(1-\gamma)} = \frac{N_{\mathrm{re}}}{1-\gamma}. \tag{C6}$$

If $\gamma = 1$, the sum can be calculated by using l'Hôpital's rule. We then get

$$N_n = \lim_{\gamma\to 1} \frac{N_{\mathrm{re}}(1-\gamma^n)}{(1-\gamma)} \left[\frac{0}{0}\right] = \lim_{\gamma\to 1} \frac{\frac{\mathrm{d}}{\mathrm{d}\gamma}N_{\mathrm{re}}(1-\gamma^n)}{\frac{\mathrm{d}}{\mathrm{d}\gamma}(1-\gamma)} = N_{\mathrm{re}}n, \tag{C7}$$

which is obvious. If $\gamma \geq 1$ and $n \to \infty$, the sum diverges to infinity since $\lim_{n\to\infty}\gamma^n = \infty$. If $\gamma > 1$ for large $n$, the sum can be expressed as

$$N_n = N_{\mathrm{re}}\gamma^n. \tag{C8}$$

**Acknowledgments**
This study was supported by the European Research Council under the European Union's Seventh Framework Programme (FP7/2007-2013)/ERC grant agreement 320839 and the Research Council of Norway under contracts 208028/F50, 216872/F50, and 223252/F50 (CoE). S. Celestin's research is supported by the French Space Agency (CNES) through a Chair of Excellence and the satellite mission TARANIS. The results in this paper are produced by simulations using GEANT4. GEANT4 version 9.6 is available for download at the official GEANT4 support web page (http://geant4.cern.ch/support/source_archive.shtml). The scripts used to set up and run the simulations can be obtained by contacting Alexander Broberg Skeltved by e-mail (Alexander.Skeltved@ift.uib.no).

Alan Rodger thanks the reviewers for their assistance in evaluating this paper.